\documentclass[usenatbib]{mn2e}
\usepackage{graphicx}
\usepackage{amssymb}
\usepackage{amsfonts}
\usepackage{amsmath}
\usepackage{aas_macros}

\bibpunct[,]{(}{)}{;}{a}{}{,}

\newcommand{\msun}{M$_\odot$}

\begin{document}

\title[dwarf spheroidal satellites from dark matter free tidal dwarf galaxy progenitors]{Dwarf spheroidal satellites of the Milky Way from dark matter free tidal dwarf galaxy progenitors: maps of orbits}
\author[R. A. Casas,  V. Arias,  K. Pe\~na Ram\'{\i}rez and P.  Kroupa]{R. A.  Casas$^1$,  V. Arias$^{1,2}$, K. Pe\~na Ram\'{\i}rez$^{1,3}$, P. Kroupa$^4$\\
		$^1$Departamento de F\'{\i}sica, Universidad Nacional de Colombia, Colombia\\
		$^2$Hamburger Sternwarte, Universit\"at Hamburg, Germany\\
		$^3$Instituto de Astrof\'{\i}sica de Canarias (IAC), Spain\\
	       	$^4$Argel\"ander Institut f\"ur Astronomie, Germany}

\date{Accepted ---. Received ---; in original form ---}

\pagerange{\pageref{firstpage}--\pageref{lastpage}} \pubyear{2012}

\maketitle
\label{firstpage}

\begin{abstract}
The long term time evolution of tidal dwarf satellite galaxies with two different initial densities orbiting a host galaxy that resembles the Milky Way has been studied using a large set of Newtonian N-Body simulations. From the simulations two maps of the orbital conditions that lead to quasi-equilibrium objects were constructed. It has been found that several orbits of the satellites allow for the existence, for about 1 Gyr or more, of out-of-equilibrium bodies with high apparent mass-to-light ratios.  Within this framework the satellites in the quasi-stable phase reproduce the observed satellite properties for about 16\% of the orbit for high density progenitors, and for about 66\% for progenitors with lower densities
An additional simulation for a single satellite with initial mass of $10^7$ \msun{} and Plummer radius of 0.15 kpc leads to remnants in the quasi-equilibrium phase that simultaneously reproduce remarkably well the observational quantities of the UFDGs of the Milky Way. This  satellite in the quasi-stable phase reproduces the observed satellite properties for about 42\% of the orbit. 
The results suggest that a fraction of the observed satellites could plausibly be galaxies without dark matter that have true M/L ratios much lower than those measured. The inflated M/L ratios arise because they are observed at the right time, along the right orbit and during the quasi-equilibrium phase of their evolution. This is a viable explanation for the high M/L ratios observed in all satellites as long as the satellites are preferentially on certain orbits and are observed at certain times. This could arise within the TDG scenario if all satellites are created at the same time along a few specific orbits that are particularly susceptible to the quasi-equilibrium phase.

\end{abstract}


\section{Introduction}
Observations suggest the existence of a large amount of dark matter in the universe, which is the dominant mass component on galactic and galaxy clusters scales. Nevertheless there is no observational evidence for dark matter on scales smaller than globular clusters.
There are several dwarf spheroidal (dSph) galaxies orbiting the Milky Way at distances ranging from tens to hundreds of kiloparsecs, with a non isotropic spatial distribution, many of them being near a plane perpendicular to the galactic plane \citep{2002ApJ...564..736P, 2010A&A...523A..32K}.  The velocity dispersions of stars in these galaxies ($\sigma \approx 10$ km/s) are similar to the ones observed in globular clusters, and have similar masses ($M_{st} \approx 10^5 M_{\odot}$) and photometric luminosities in the V band ($L \approx10^5 L{\odot}$), but the dSphs are approximately a hundred times more extended than globular clusters ($R \approx 300$ pc). Some dSph satellites present a large mass to light ratio, implying them to be completely dominated by dark matter. 

In recent years several spheroidal ultra faint dwarf galaxies (UFDG) have been found in the halo of the Milky Way \citep{2005AJ....129.2692W,2005ApJ...626L..85W,2006ApJ...653L..29S,2006ApJ...650L..41Z,2006ApJ...643L.103Z, 2006ApJ...647L.111B, 2007ApJ...654..897B, 2007ApJ...662L..83W, 2008ApJ...686L..83B} using the SDSS-DR6 \citep{2000AJ....120.1579Y, 2008ApJS..175..297A} catalogue. These satellite galaxies exhibit extremely low stellar densities. Their half light radii are relatively similar to those of the dwarf spheroidal satellite galaxies of the Milky Way known prior to SDSS \citep{2000AJ....120.1579Y} (dSphs). The kinematics of the eight UFDGs: Ursa Major II, Leo T, Ursa major I, Leo IV, Coma Berenices, Canes Venatici II, Canes Venatici I and Hercules have been measured by \citet{2007ApJ...670..313S}. The reported values for central velocity dispersions of five of the UFDGs are lower than those of the dSphs, while four of the UFDGs present central velocity dispersions similar to the lower end of the corresponding values for the dSphs. On the other hand the UFDGs exhibit  lower values for the central surface brightness in comparison with the dSphs with no intersection at all. While the dSphs have mass-to-light ratios up to 100, the UFDGs have mass-to-light-ratios up to $\sim$1800 \citep{2007ApJ...670..313S}, assuming the satellites are in virial equilibrium.

Generally it is assumed that these galaxies are cold dark matter dominated, due to their high mass-to-light-ratios. Nevertheless, one possible alternative explanation, within the standard cosmological framework, for the large mass to light ratios of dSphs and UFDGs is that they are objects out of virial equilibrium and that they may not be spherical but with non isotropic velocity dispersions, i.e., the dynamical mass to light ratios may not be real \citep{1998ApJ...498..143K}. In that case the masses of the dwarf satellites would be overestimated since the mass-to-light ratios are obtained from an in-equilibrium assumption.
\citet{1997NewA....2..139K} performed simulations of the evolution of an initially spherical satellite with 300.000 particles and a mass of $10^7\ \mathrm{M}_{\odot}$ on different orbits. He used an extended dark matter halo with circular velocity of 200 km/s for the host galaxy and found that unbound remnant quasi-stable systems without dark matter, resembling dSph galaxies, may exist. A remarkable result of this work was the prediction of a satellite galaxy that nearly exactly matches the Hercules dSph discovered ten years later \citep{2010A&A...523A..32K}. On the other hand  the  infall of dark matter satellites from a filament onto the Milky Way has been ruled out \citep{2011MNRAS.tmp.1233A}.

\citet{2002ApJ...566..838K} showed that if a dSph satellite of the Milky Way has no dark matter and is out of equilibrium there must be a significant spread in the magnitude of the horizontal-branch (HB) stars that is correlated with sky position and velocity. They  proposed this feature as a test for the tidal models of dSph galaxies. The apparent magnitude range of the widened HB stars obtained by \citet{2002ApJ...566..838K} for an out-of-equilibrium satellite similar to Sat-1 is $\approx 1$ mag, which is near the value of $0.8$ we have used to cut the projected satellites along the line-of-sight in our simulations. Nevertheless, this magnitude limiting value should be tuned in the case of a study of a particular real galaxy. 
\citet{2003ApJ...589..798K} applied to Draco the observational test proposed
by  \citet{2002ApJ...566..838K}. They found a narrow HB depth of about 0.13 
mag to be almost constant with increasing observation field size and therefore they 
concluded that Draco cannot be the remnant of a tidally disrupted satellite
but is probably dark matter--dominated.  Nevertheless, the ratio of the
line-of-sight depth for Draco, corresponding to the HB depth of about 0.13 
mag and a distance of $80$ kpc,  is about 4800 pc, which is about 27 times
larger than the half-light-radius of Draco ($180$ pc). Thus, even with a
narrow HB depth, as reported by  \citet{2003ApJ...589..798K}, Draco can still
be a very elongated object along the line-of-sight and therefore a tidally disrupted object.

Recently, \citet{Pawlowski:2012} have discovered a vast polar structure (VPOS) around the MW, which extends from about 10kpc to at least 250 kpc and contains stellar and gaseous streams, all young halo globular clusters and all satellite galaxies. This VPOS is a highly correlated structure in phase-space and falsifies the hypothesis that the satellite galaxies are in DM halos with individual formation histories with very high significance. The VPOS is consistent with the only currently known alternative, namely that it is the remnant of an ancient tidal arm that was pulled out of the young MW about 10--11 Gyr ago when the bulge of the MW formed during an encounter with another galaxy \citep{2011A&A...532A.118P}. Furthermore \citet{2012PASA...xxx..xxK} has shown that the dark matter standard model of cosmology is incompatible with a large set of galactic and  extragalactic observations. These results motivate us here to reconsider the Ansatz by Kroupa (1997) that the satellite galaxies may be DM free ancient tidal-dwarf galaxies. 

In this paper we present a study of the orbital conditions for an initially in-equilibrium satellite without dark matter orbiting a galaxy that resembles the Milky Way, using a large set of numerical Newtonian N-body simulations. The host galaxy is simulated through a three component rigid potential, while the initial satellite corresponds to a Plummer sphere with $10^6$ particles and masses of $10^7$ and $10^8$  M$_{\odot}$ respectively.  The simulated orbits cover a wide range of eccentricities and apocentric distances. We show the dynamical evolution of the satellites and some of their observational properties as they would appear to an observer on Earth. Furthermore, we present two maps for the parameters of the orbits that allow the formation of out-of-equilibrium remnants with high apparent mass-to-light ratios that could appear to an observer as dark matter dominated systems. The observational properties of the quasi-equilibrium objects are compared with the corresponding quantities reported in the literature for real dSphs and UFDGs of the Milky Way. 

\section{Models}
\subsection{Model for the Galaxy}
The main galaxy is modeled using a rigid potential with three components:
a Miyamoto--Nagai potential for the disk
\begin{equation}
\phi_{disk}(R,z)=-\dfrac{GM_d}{\sqrt{R^2+(a+\sqrt{z^2+b^2})^2}},
\end{equation}
a Hernquist spherical potential for the bulge
\begin{equation}
  \phi_{sph}(R)=-\frac{GM_b}{R+R_H},
\end{equation}
and a Logarithmic potential for the dark matter halo
\begin{equation}
  \phi_{halo}(R,z)=\frac{1}{2}v_0^2 ln\left[
    R_c^2+R^2+\frac{z^2}{q_{\phi}^2}\right],
\end{equation}

\noindent where $R^2 = x^2 + y^2$, $M_d =10^{11}$ M$_{\odot}$ is the mass of the disk, $a = 6.5$ kpc, $b=0.26$ kpc are constants to determine the geometry of the disk; $M_b=3.4\times 10^{10}\ M_{\odot}$ is the mass of the bulge  and $R_H=0.7$ kpc is a concentration parameter; $R_c= 12$ kpc  and $v_0=128$ km/s are constants to define the density distribution in the halo, $q_{\phi} = 1$.  These potentials and the corresponding parameters have been chosen since they resemble the potential of the Milky Way \citep{1995ApJ...451..598J}.  

\subsection{Model for the satellite}
The initial object is modelled as a Plummer sphere with $10^6$ particles,  a Plummer radius of $b = 0.3$ kpc and a cutoff radius of $1.5$ kpc,
\begin{equation}\label{ Plummer}
\rho(r)= \left(\frac{3M}{4\pi
  b^3}\right)\left(1+\frac{r^2}{b^2}\right)^{-\frac52}
\end{equation}

\noindent and
\begin{equation}\label{ Plummer-pot}
\phi_{Pl}(r)= \frac{GM}{\sqrt{r^2 + b^2}}.
\end{equation}

\begin{figure}
	\includegraphics[width=\linewidth,clip]{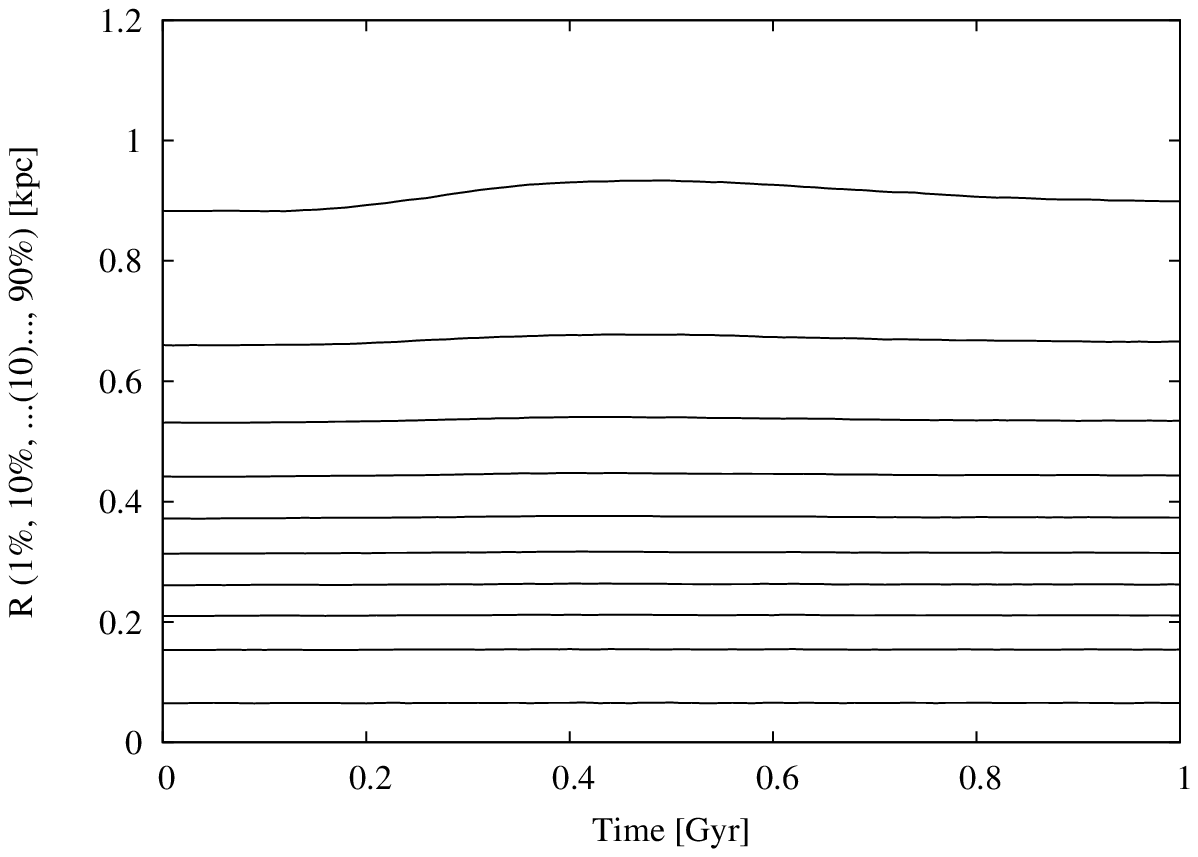}\\
	\includegraphics[width=\linewidth,clip]{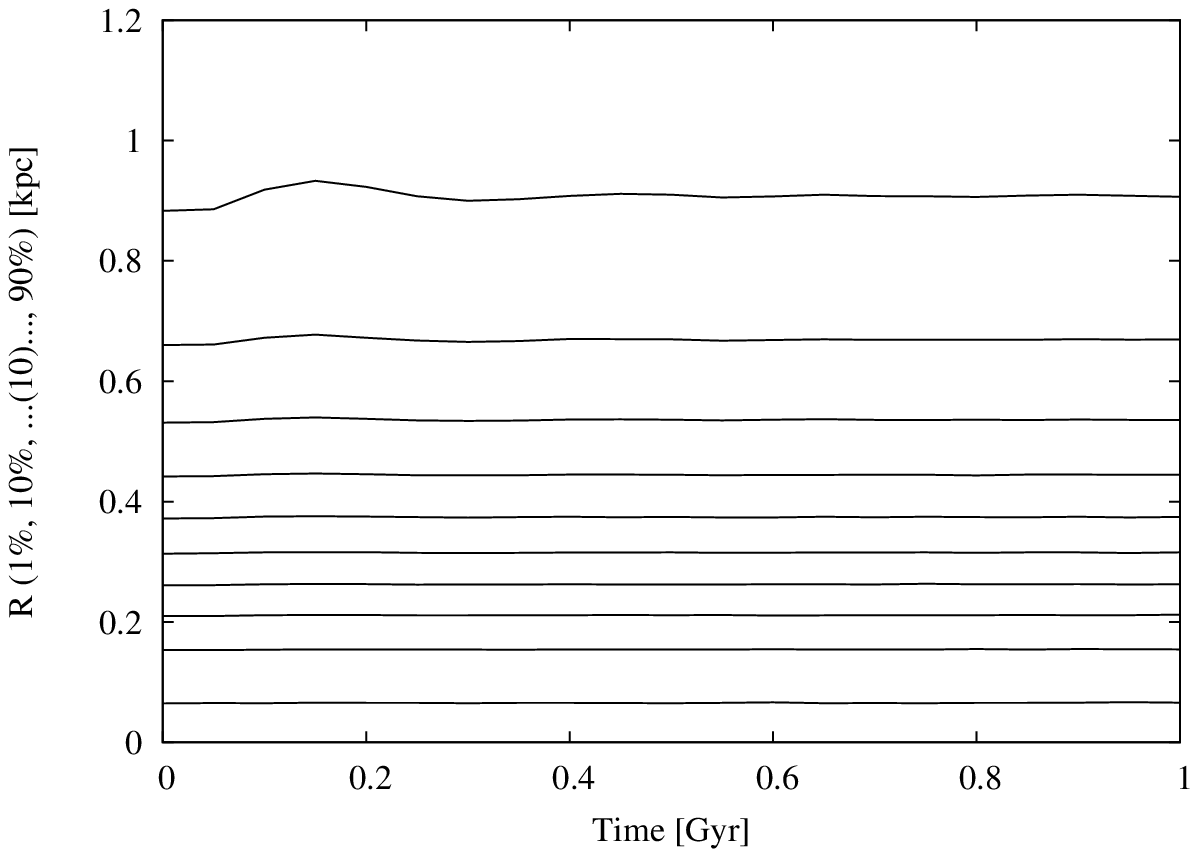}
 \caption{Time evolution during virialisation of the Lagrange radii of the Plummer spheres Sat-1 (top) and Sat-2 (bottom). The radial density profiles of the objects prior and after virialisation, along with a fit to  Plummer profiles, are shown in Figure \ref{fig:virialisation-profiles}.}
	\label{fig:virialisation}
\end{figure}

\begin{figure}
\includegraphics[width=\linewidth,clip]{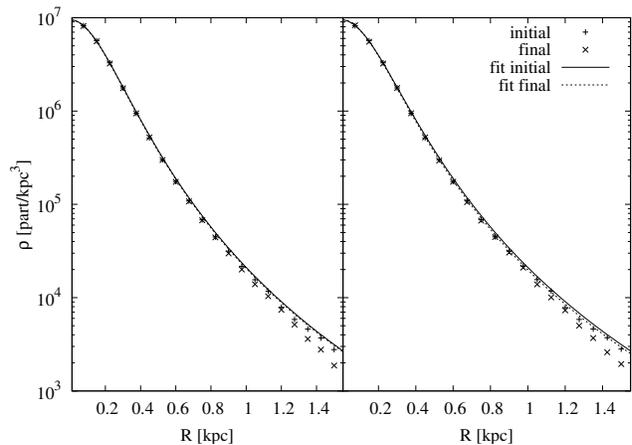}
 \caption{Radial density profiles of the objects prior and after virialisation for Sat-1 (left) and Sat-2 (right). The corresponding Plummer profile fits are shown as lines.}
	\label{fig:virialisation-profiles}
\end{figure}

Two masses of the initial object are used: $10^7$ M$_{\odot}$ (Sat-1) and $10^8$ M$_{\odot}$ (Sat-2). The satellites have been constructed using the algorithm proposed by \citet{1974A&A....37..183A}; See also \citet{Kroupa:2008Lect}. Since the discretized object is not in complete equilibrium, it is necessary to relax the system of particles under its self gravity. Therefore, after construction the evolution of the isolated object is simulated for a couple of Gyr using Gadget-2 \citep[]{2005MNRAS.364.1105S} until it reaches equilibrium. The time evolution during virialisation of the Lagrange radii of the two Plummer spheres (Sat-1 and Sat-2) is shown in Figure \ref{fig:virialisation}. The density profiles of the objects after they have attained dynamical equilibrium are fitted to a Plummer profile (Figure \ref{fig:virialisation-profiles}), verifying that the objects after virialisation are still  Plummer spheres with practically the same initial parameters. Note that the initial density of Sat-1 is 0.09 \msun{}$/pc^3$ and of Sat-2 is 0.9 \msun{}$/pc^3$.


\section{N-body simulations}

The orbits that a satellite can describe around the Milky Way are completely determined by their eccentricity and apocentric distance. In this study, the initially spherical virialised satellite is introduced into the gravitational potential of the Milky Way at a given  apocentric distance in the plane of the Galactic disk, with velocity perpendicular to the plane of the disk. The speed of the satellite depends on the eccentricity of the orbit and on the potential of the host galaxy.

The evolution of the satellite orbiting the Milky Way has been simulated for several tens of different orbits around the Galaxy, covering a grid in apocentric distances and eccentricities. The eccentricities were simulated in steps of 0.1, while the apocentric distances have a step of 10 kpc. The simulations were performed using the tree-code feature of the Gadget-2 N-body simulations code, modified to include the rigid potential of the Milky Way.  For the simulations a softening length of 7 pc and the relative cell-opening criterion have been used. The relative cell-opening criterion limits the absolute truncation error of the multipole expansion for every particle-cell interaction and delivers higher force accuracy compared to the Barnes-Hut criterion \citep[]{2005MNRAS.364.1105S}. Each simulation has been run for 12 Gyr, saving a snapshot of the positions and velocities of the particles in the satellite every 0.1 Gyr. These snapshots are used to follow the evolution of some quantities of the satellites like their Lagrange radii, central surface brightness, half light radius, central line-of-sight velocity dispersion and the mass to light ratio. The last four quantities have been obtained projecting the remnant object as it would appear to an observer on Earth.

\section{Dynamical evolution of the satellites}

\begin{figure*}
	\includegraphics[width=0.49\linewidth,clip]{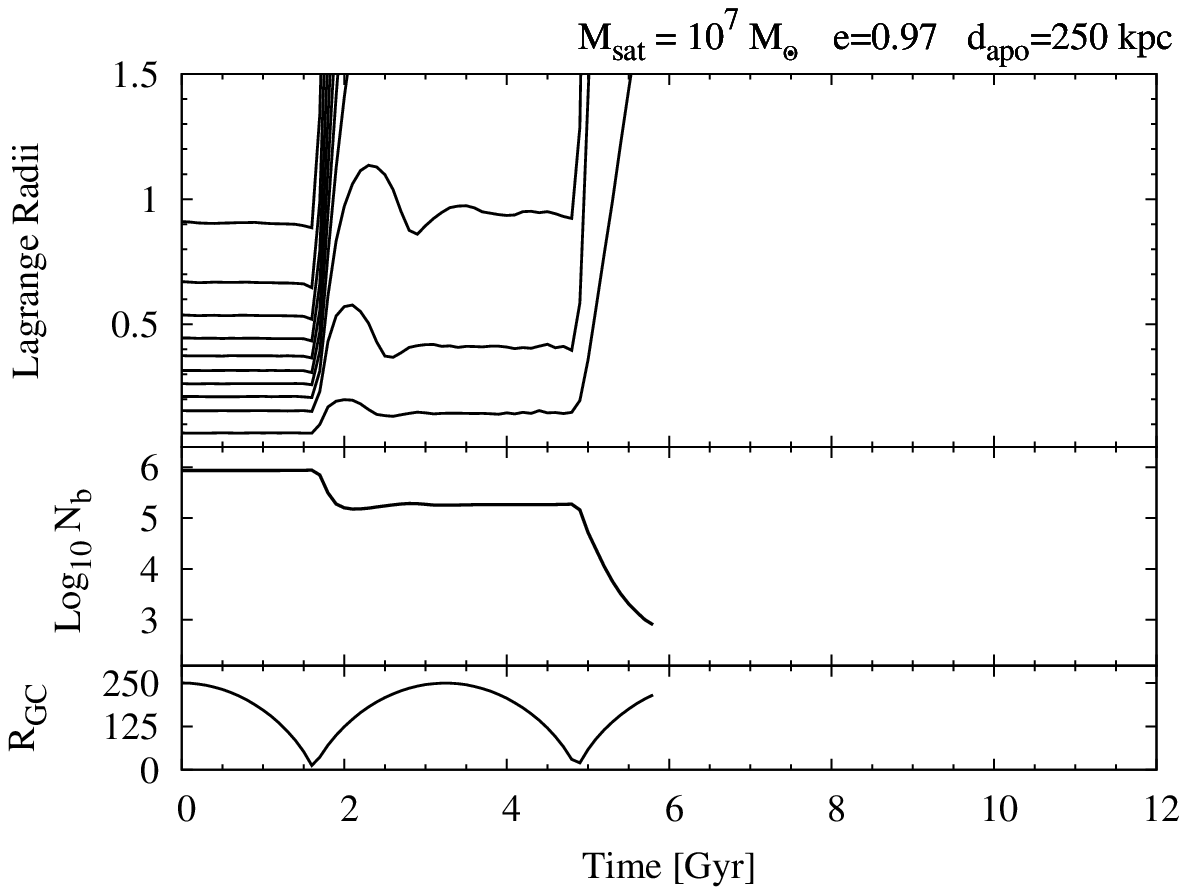}
	\includegraphics[width=0.49\linewidth,clip]{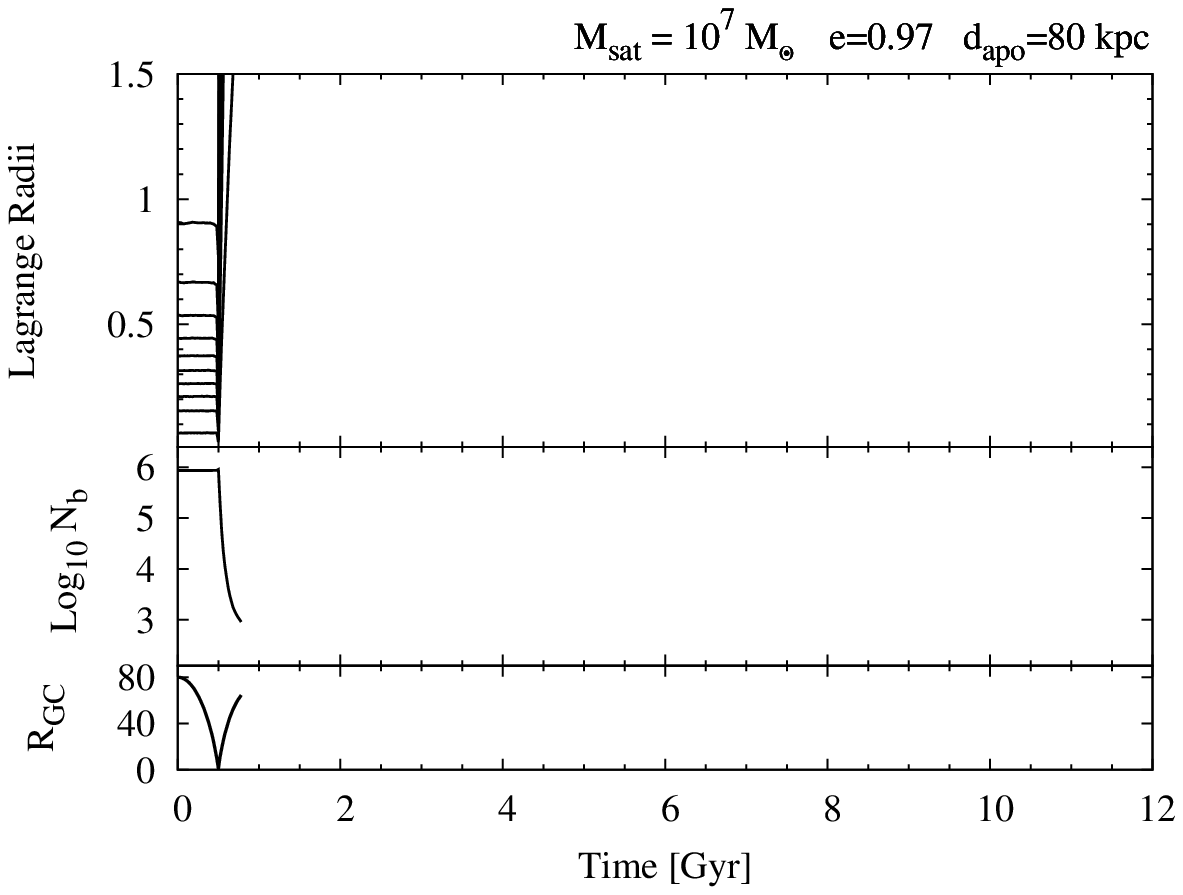}\\
	\includegraphics[width=0.49\linewidth,clip]{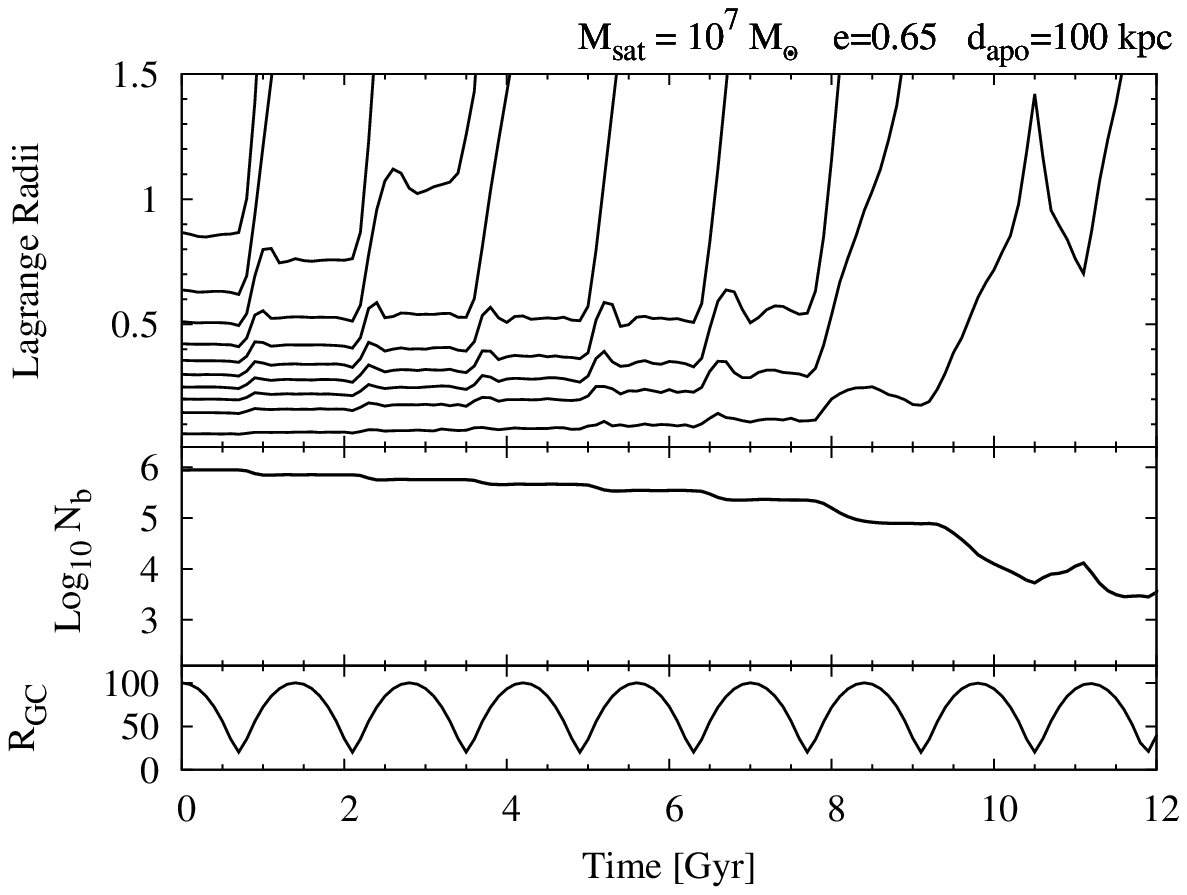}
	\includegraphics[width=0.49\linewidth,clip]{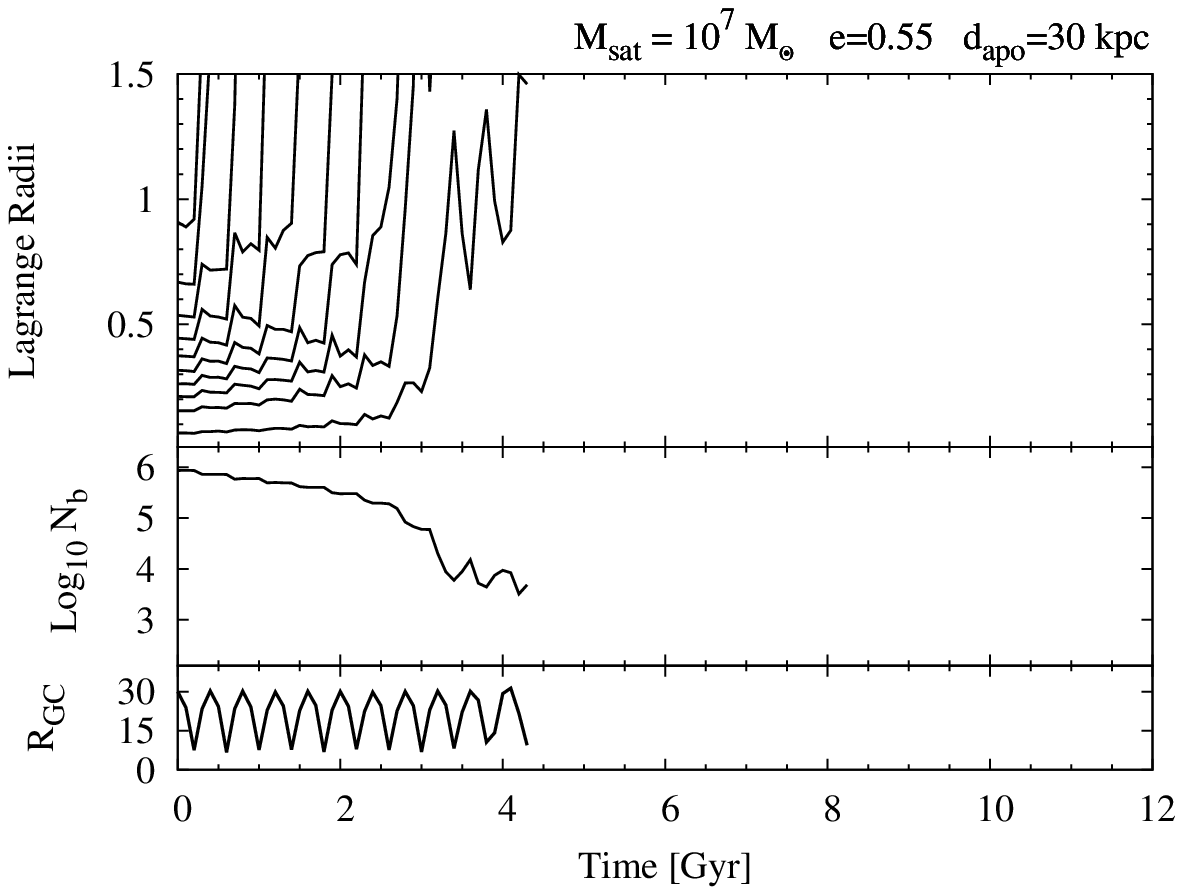}\\
        \includegraphics[width=0.49\linewidth,clip]{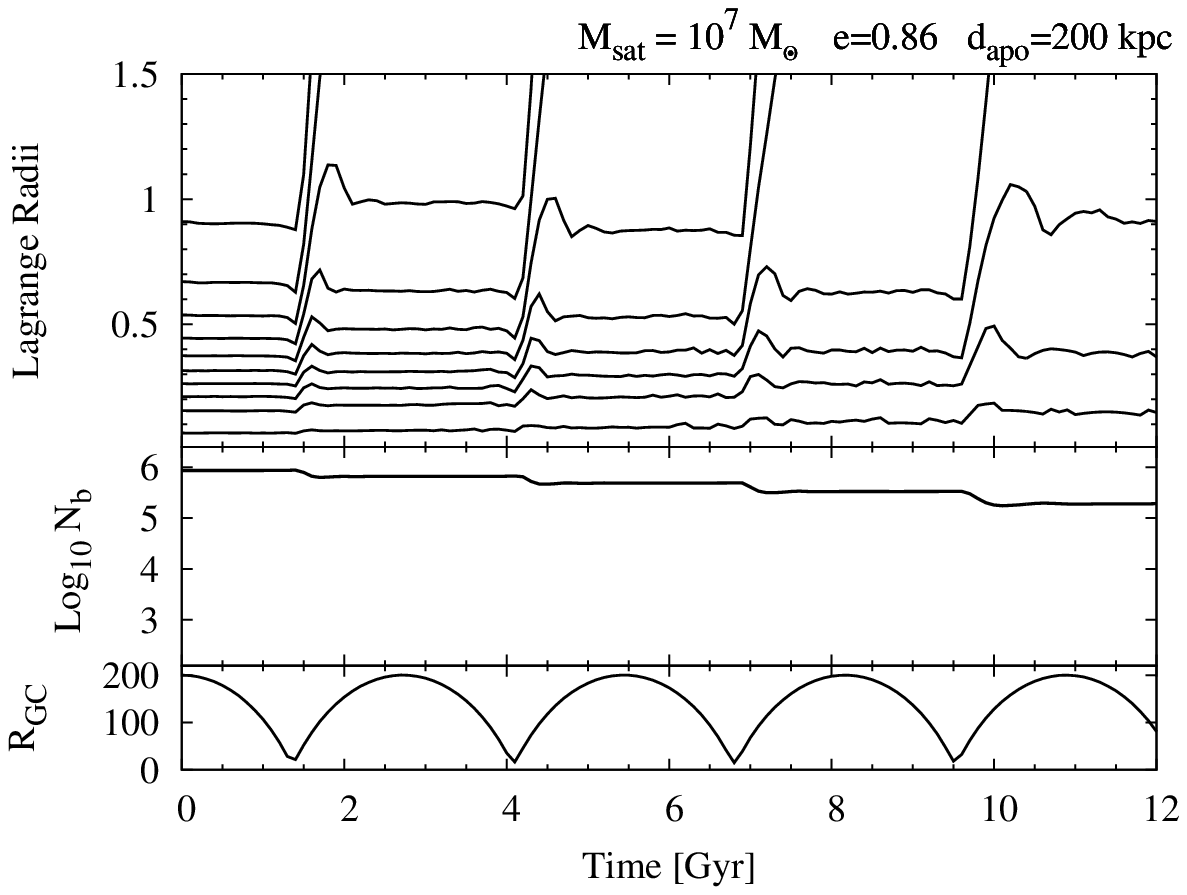}
        \includegraphics[width=0.49\linewidth,clip]{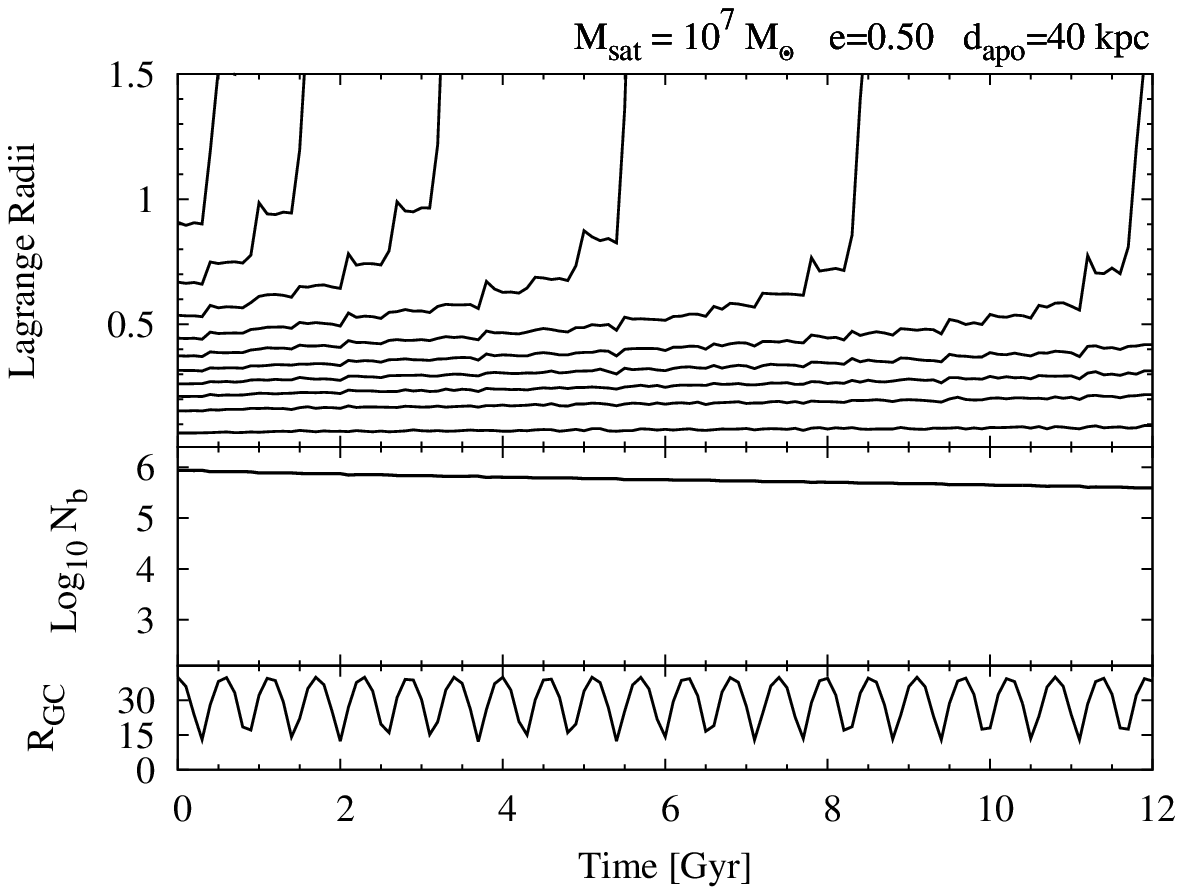}
 \caption{Evolution of six satellites in six plots. In each plot the top panel shows the Lagrange radii containing (bottom to top) 1\%, 10\%, 20\%, ... , 90\% of the initial mass of six Sat-1 type (left) and  Sat-2 type (right) satellites. The middle panel shows the evolution of the number of particles inside a core sphere of 0.8 kpc and the bottom panel shows the Galactocentric distance of the density maximum. The parameters of the orbit are given in the upper right corner. The Lagrange radii and Galactocentric distances are given in kpc.}
\label{fig:lagrangeradii}
\end{figure*}

In Figure \ref{fig:lagrangeradii} we show the Lagrange radii containing 1, 10, 20, 30, 40, 50, 60, 70, 80 and 90 percent of the initial mass of the corresponding satellite. These Lagrange radii are useful to describe the overall evolution of the satellite and to see how it is depopulated by the gravitational interaction with the host galaxy. In the Figure we present the time evolution of six satellites as they orbit the host galaxy. There are three possible tracks of evolution: satellites that loose almost all their initial mass very rapidly (top panels in Figure \ref{fig:lagrangeradii}); satellites that keep more than 10\% of their initial mass for a Hubble time (bottom panels); and satellites that loose more than the 90\% of their initial mass in a Hubble time, but still maintain around 1\% of their initial mass (middle panels). These satellites are the ones we are most interested in, since they are disrupted objects, i.e. out-of-equilibrium but quasi-stable remnants.

The number of particles ($N_b$) inside a radius $r < 0.8$ kpc centred on the maximum of density of the object is used as an approximation to the number of dynamically bound particles of the satellite \citep{1997NewA....2..139K}. Although this count might be contaminated by particles instantly inside the sphere, it is expected that the effect of this contamination should be negligible due to the high speeds of those particles. In the plots shown in Figure \ref{fig:lagrangeradii} we present the number of particles inside the core for the  corresponding satellites. It can be observed that the number of core particles decreases very slowly while the satellite is slowly depopulated until the perigalactic passage when about 90\% of the initial mass of the satellite has been subtracted. After that, the number of core particles decreases rapidly due to the low binding energy of the object, which is now in a quasi-stable phase. The number of core particles is a good indicator for the dynamical state of the satellite.

In general, the depopulation of a satellite with initial density like Sat-2 takes more orbital periods than in the case of the less dense satellite Sat-1. Thus, if a satellite with an initial density like Sat-1 in a given orbit is depopulated so that it reaches the quasi-stable phase within a Hubble time, an initially denser satellite (like Sat-2) will spend a longer time in its in-equilibrium phase and, depending on the orbit, it might not reach a quasi-stable phase within a Hubble time. That also means that satellites initially much less dense than Sat-1 are depopulated faster than Sat-1 and therefore a satellite with an initial mass of $10^6\ \mathrm{M}_\odot$ with otherwise the same parameters as Sat-1 would reach the quasi-equilibrium phase within a Hubble time on orbits that Sat-1 would never show the quasi-equilibrium phase on. Also, it will be rapidly destroyed on some orbits where Sat-1 still reaches the quasi-equilibrium phase. Satellites with initial masses lower than $10^7\ \mathrm{M}_\odot$ but with the same density (i.e., a smaller Plummer radius) as Sat-1 or Sat-2 would be depopulated in the same way as them and will exhibit the same time evolution and quasi-equilibrium phase as Sat-1 and Sat-2 respectively, but with some of their observational properties scaled appropriately. 

\section{Initial conditions maps}

\begin{figure}
\includegraphics[width=\linewidth,clip]{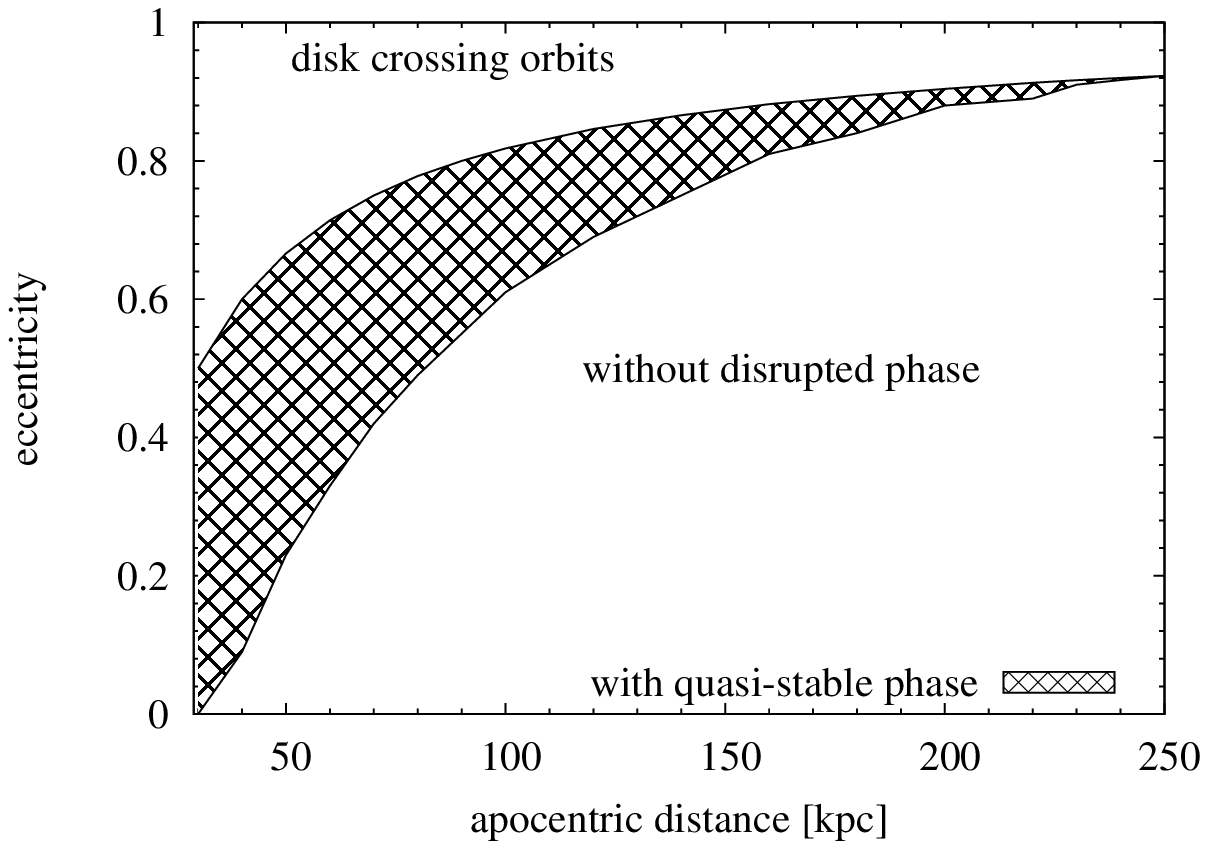}\\
\includegraphics[width=\linewidth,clip]{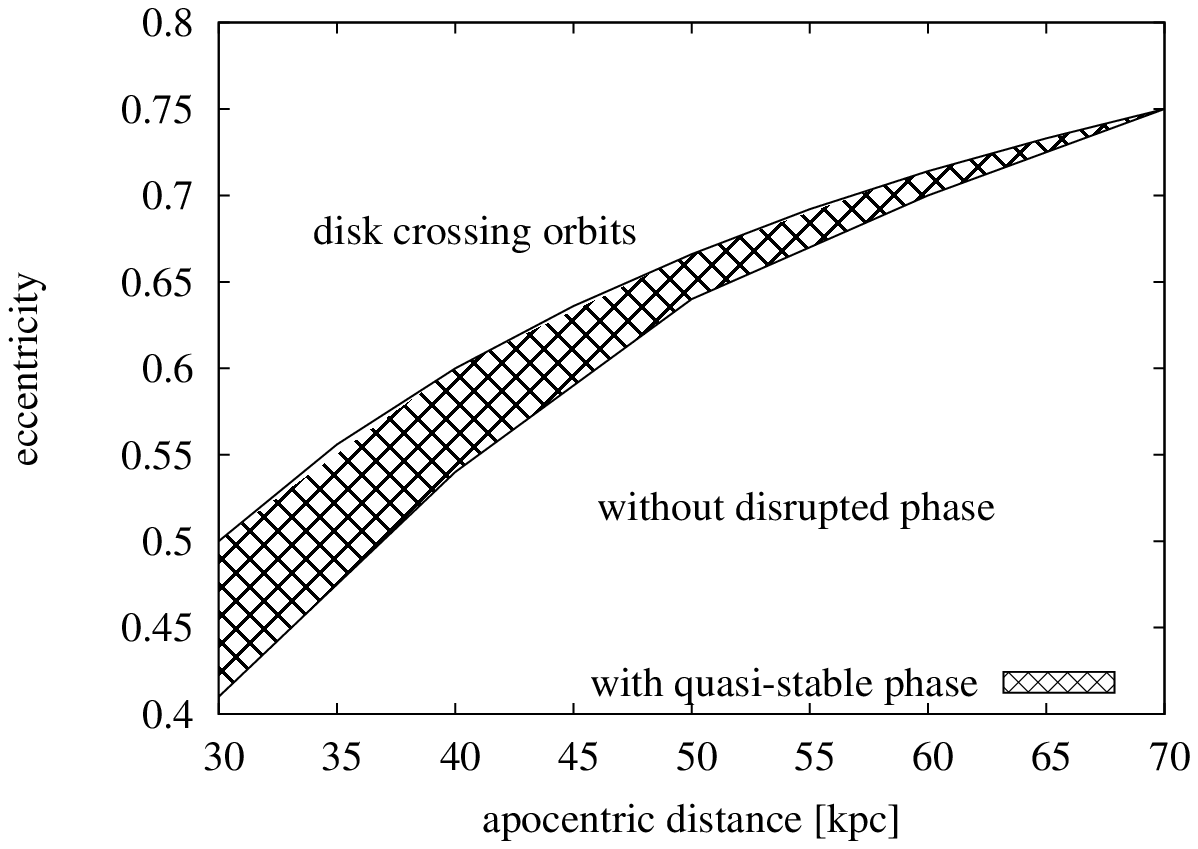}
\caption{Maps of the orbital parameters that lead to three different fates of the satellites. Top: initially less dense satellites. Bottom:  initially more dense satellites. Shaded regions correspond to orbits that lead to quasi-stable satellites in Newtonian dynamics.}
\label{fig:maps}
\end{figure}

As the satellite orbits the host galaxy, it looses mass due to tidal forces that act on its stars. The rate at which the initial satellite is depopulated depends on its concentration, i.e., on its mass and  Plummer radius, and on the eccentricity and apocentric distance of its orbit. From the suite of simulations performed we constructed two maps  that show the fate of a satellite on a given orbit. 
We did not take into account satellites on orbits with perigalactic distances smaller than 10 kpc in order to avoid satellites crossing the Milky Way disk, since the rigid potential approximation we use might not be accurate to simulate the encounters between the disk and the satellite.
The two initial satellites used in the simulations are similar objects with different densities, since their densities obey Plummer profiles with the same Plummer radius and cut-off radius and they differ only in the total mass. The initial satellite with a mass of $10^8$ M$_{\odot}$ (Sat-2) is 10 times denser than the initial satellite with a mass of $10^7$ M$_{\odot}$ (Sat-1). 
 In Figure \ref{fig:maps} the maps corresponding to the initial less dense (Sat-1) and more dense (Sat-2)satellites are presented. For the construction of the maps we followed the evolution of the Lagrange radii and number of core particles of each simulated satellite. If the Lagrange radius containing 10\% of the initial mass is larger than the initial cutoff radius (1.5 kpc) and the Lagrange radii containing 1\% of the initial mass is smaller than the initial cutoff radius, then this object is classified as a quasi-stable object. With this condition in mind it appears that there are three possible evolution tracks for the objects: they can survive for a Hubble time without reaching a quasi-stable phase, they can reach a quasi-stable phase and stay there for more than 1 Gyr or they can be depopulated so fast that the quasi-stable phase lasts less than 1 Gyr and the satellite is rapidly destroyed. Although it would be possible to detect rapidly destroyed satellites in their quasi-stable phase, it is not very likely to happen due to their short life-time in that phase. 

From Figure \ref{fig:maps} it is also noticeable that high density satellites and apocentric distance larger than 60 kpc cannot have a quasi-equilibrium phase within a Hubble time, while this maximum apocentric distance is larger for the less dense satellites (Sat-1). 
The  Plummer radius has been kept the same for both sets of simulations. Objects with larger  Plummer radii are less compact than objects with a smaller  Plummer radius and the same mass. Thus, increasing the  Plummer  radius would make our satellites more easily depopulated lowering the lower curve in Figure \ref{fig:maps} enlarging the eccentricity -- apocentric distance region that lead to quasi-stable objects within a Hubble time.

In Figure \ref{fig:maps} it can be observed that for less dense progenitors the set of orbits that could lead to satellites in the quasi-equilibrium phase is larger than for more dense objects. Therefore, as the initial density of the satellite decreases the probability that it can evolve into an object in the quasi-equilibrium phase increases. This leads us to the conclusion that initially much denser objects than Sat-2 are not likely to evolve into a quasi-stable phase within a Hubble time. In particular we verified that a satellite with an initial mass of $10^8$ M$_{\odot}$ and a Plummer radius of 0.15 kpc can not evolve into a quasi-equilibrium phase within a Hubble time on any orbit. We did not simulate more massive objects since their masses and low mass loss rate could imply that dynamical friction might be non negligible. On the other hand, objects less dense than Sat-1 are very likely to evolve into a quasi-stable phase within a Hubble time. Therefore, some of the dSphs and UFDGs of the Milky Way may correspond to remnants of initially low density spheroidal systems that are in a phase of dynamical quasi-equilibrium.  

The nine dSph satellites of the Milky Way are located at a range of distances from the Galactic centre spanning from 24 $\pm$ 2 kpc for Sagittarius to 250 $\pm 30$ kpc  for Leo I. The arithmetic mean of these distances corresponds to 115 kpc and only 2 galaxies (Leo I and Leo II) are located beyond 140 kpc from the Galactic centre. The UFDGs of the Milky Way: UMa II, Wil 1, CBe, Boo II, Boo, UMa, Her, CVn II, Leo IV, Leo V and CVn are located in a range that spans from 36.5 kpc to 220 kpc, with no objects between 60 and 100 kpc. 

The probability that a satellite could experience a quasi-equilibrium phase lasting more than 1 Gyr as it disrupts along any orbit can be computed as the ratio between the areas of the shaded region and the region under the disk-crossing line in the maps of Figure \ref{fig:maps}. This probability is 20\% for the case of progenitors with initial low density and with galactocentric distances between 30 and 240 kpc; if the Galactocentric distance range is reduced to 30 -- 150 kpc, the probability increases to 35\%. In the case of the more dense progenitors this probability is 15\% at Galactocentric distances between 30 and 60 kpc and zero for distances larger than 60 kpc. The probabilities are larger still for less-massive satellites.

On the other hand, if the MW satellites have orbital parameters well outside the shaded region in Figure 4 then Newtonian dark-matter free Sat-1 and Sat-2 type solutions would become unlikely.

\section{Projected satellites}
In the last section we have shown that there are some satellites (i.e., orbits + initial mass) that present a quasi-stable phase. Now, we present the evolution of some features of a few satellites that have shown a  quasi-stable phase that lasts $\approx$1 Gyr or larger. We projected our simulated satellites as if they were seen by an observer on Earth. The projected satellite is then used to estimate some observational quantities like central surface brightness, half light radii, central line-of-sight velocity dispersion and the mass-to-light ratio. 

The observational quantities are computed using the projected particles, as observed from Earth's position, contained in a circle of radius 1.5 kpc with centre on the density maximum of the satellite. Nevertheless, some particles inside the projected circle might lie too far away from the position of the density maximum of the satellite and have to be discarded. Thus only particles lying in a given observational magnitude range are taken into account
\begin{equation}
M_{cod} - \frac{\Delta M}{2} \leq M \leq M_{cod} + \frac{\Delta M}{2}.
\end{equation}
This magnitude range can be converted into a distance interval 
\begin{equation}
d_{cod}\times 10^{-\Delta M/10} \leq d_p \leq d_{cod}\times 10^{\Delta M/10},
\end{equation}
where $d_{cod}$ is the distance from the position of the density maximum (centre of density) to the observer and $d_p$ is the distance from the particle to the observer. $\Delta M$ is the range of observable magnitudes. We used  $\Delta M = 0.8$ mag, following \citet{1997NewA....2..139K}.
For the computation of the model line-of-sight velocity dispersion the bi-weight scale estimator has been used \citep[]{1990AJ....100...32B}, with a tuning constant of 6.0 for both scale and location estimators. The bi-weight scale and location are robust tools to outlying velocity data. 

\begin{figure*}
    \centering
    (a)\\
    \includegraphics[width=0.49\linewidth]{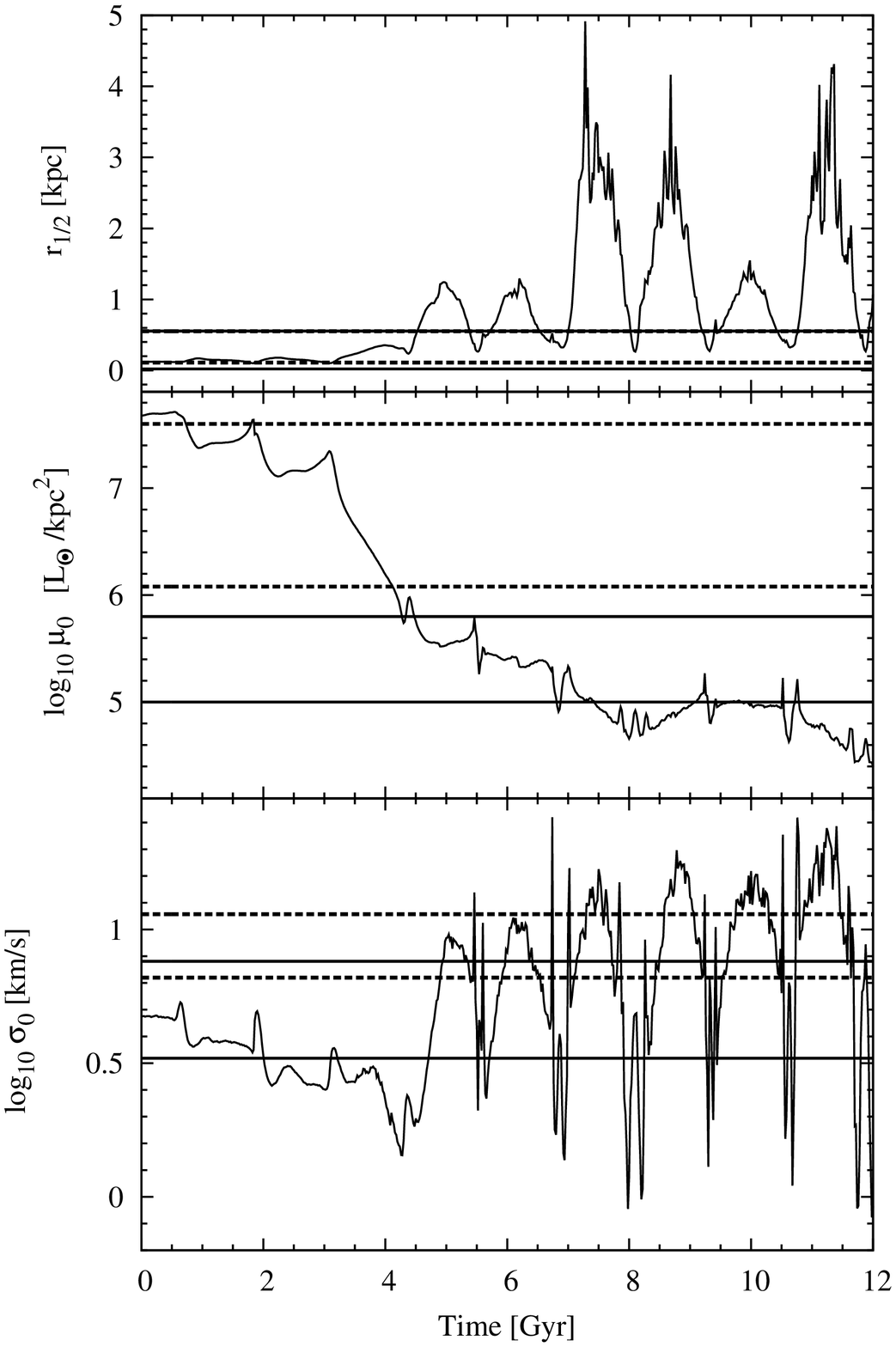}
    \includegraphics[width=0.49\linewidth,clip=yes]{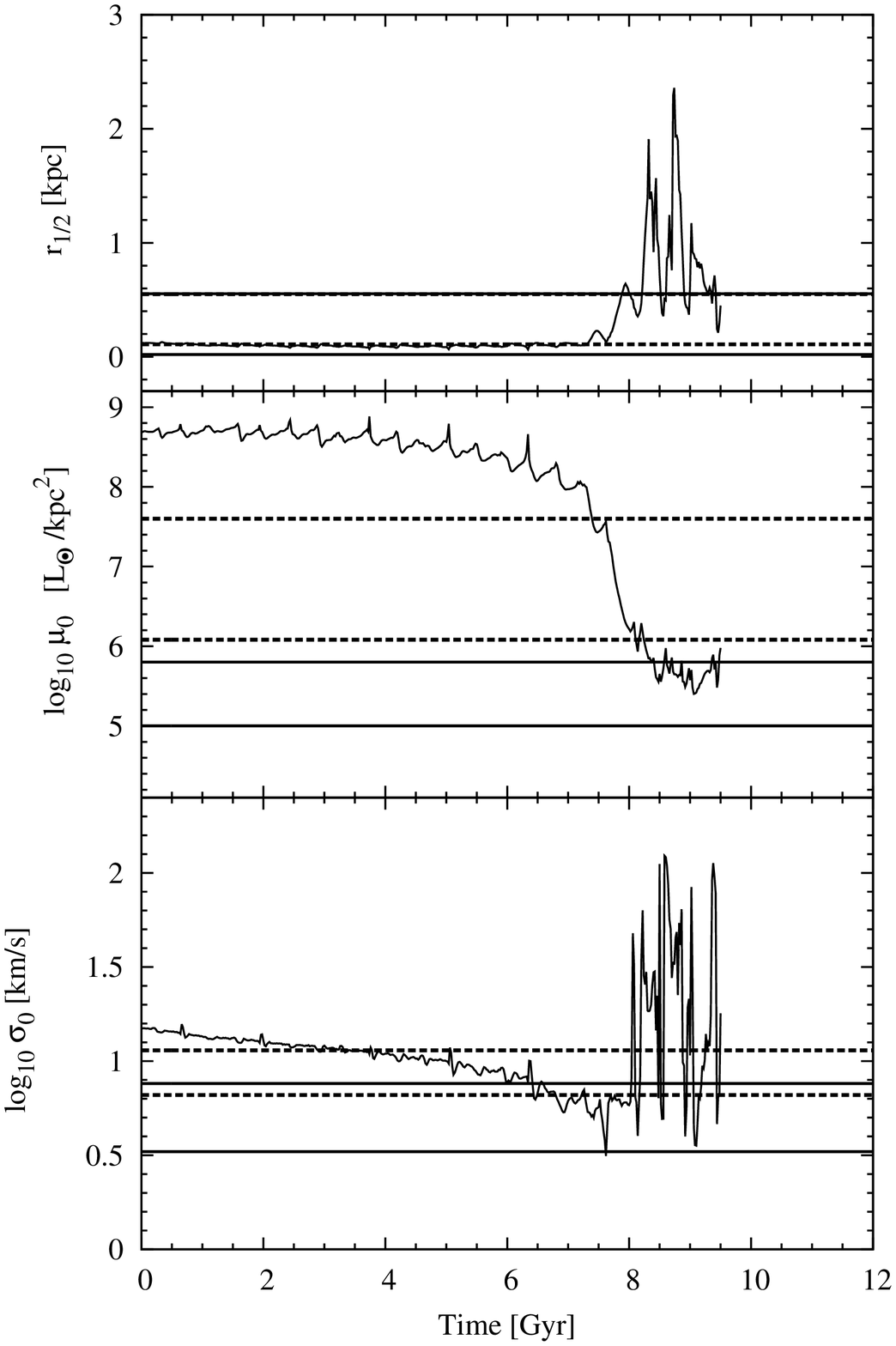}\\
    (b)\\
    \includegraphics[width=0.49\linewidth,clip=yes]{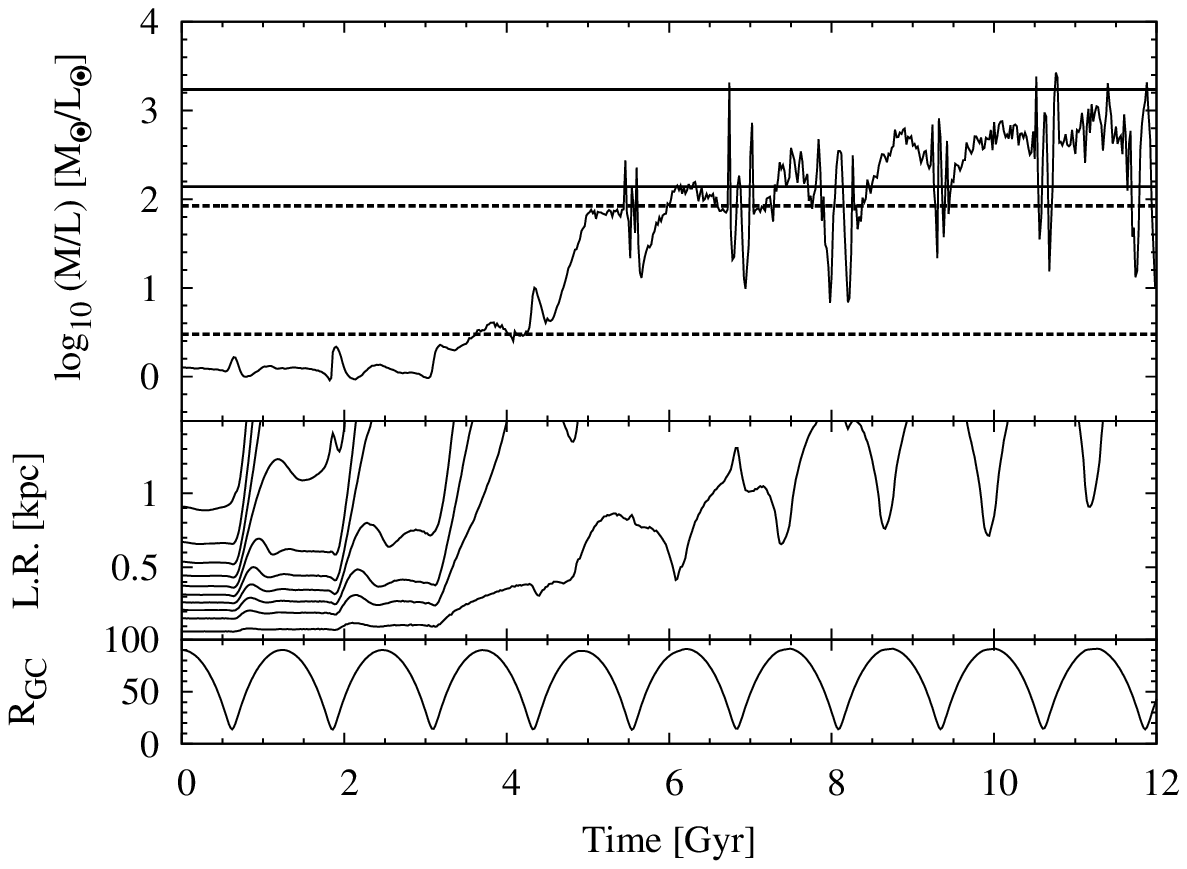}
    \includegraphics[width=0.49\linewidth,clip=yes]{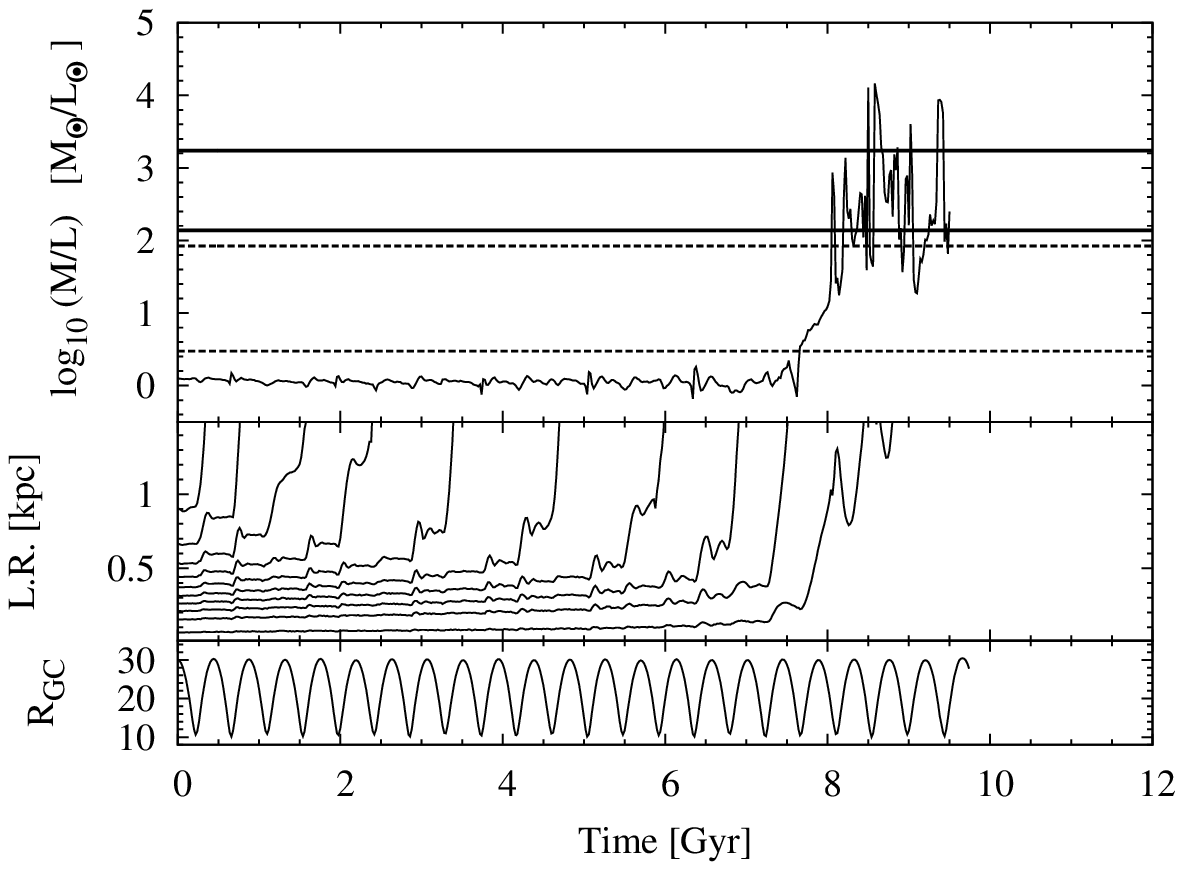}
   \caption{Observational quantities of two representative satellites that exhibit a quasi-stable phase: Sat-1 (left) and Sat-2 (right). (a) Half light radius, Central surface brightness and central velocity dispersion along the line-of- sight. (b) Mass to light ratio, along with the corresponding Lagrange radii and Galactocentric distance of the centre of density of the objects (in kpc). The horizontal lines show the range of observational data reported in the literature for classical dSph galaxies (dashed) and the new UFDGs (solid) of the Milky Way.} 
    \label{fig:psats}
\end{figure*}

In Figure \ref{fig:psats}(a) we present the time evolution of the half light radius (top), the central surface brightness (middle) and the central velocity dispersion along the line-of-sight (bottom) that would be measured by an observer on Earth for two satellites that present a quasi-stable phase. The left panel corresponds to  Sat-1 on an orbit with apocentric distance of 90 kpc and eccentricity of 0.73, while the right panel corresponds to Sat-2 on an orbit with apocentric distance of 30 kpc and eccentricity of 0.45.

 It can be noticed that the half light radius remains almost constant while the satellite is smoothly depopulated. After disruption time the half light radius grows rapidly reaching values near to those reported in the literature for dSphs and UFDGs. The half light radius shows large oscillations during the quasi-stable phase of the satellite, anti-correlated with oscillations in the Lagrange radius containing 1\% of the initial mass.
  
 For both initial densities of the satellites it can be noticed that at the time the satellite enters into the quasi-stable phase (i.e., has lost about 90\% of its initial mass) the value of the central surface brightness decreases rapidly for a short time interval and then continues to slowly decrease further as in the stable phase.  The remnant thus achieves a quasi-stable state. The simulated values of the central surface brightness during the quasi-stable phase of the evolution of both satellites partially cover the ranges of values reported in the literature for the classical dSphs and the new UFDGs of the Milky Way.

In addition, the value of the central velocity dispersion along the line-of-sight decreases smoothly during the stable phase and starts to rise as soon as the satellites go into their quasi-stable phase. In the last phase the value of  the central velocity dispersion also shows large oscillations that are anti-correlated with the oscillations of the 1\% Lagrange radius.

\subsection{Mass to light ratio}

The model mass-to-light ratio of the projected satellite is obtained using the core fitting formula \citep{1972ApJ...175..627R,1986AJ.....92...72R}:
\begin{equation}
\left( \frac{M}{L} \right)_{obs} = \frac{9}{2\pi\ G} \frac{\sigma_0^2}{\mu_0 r_{1/2}},
\label{eq:m2l}
\end{equation}
where $G$ is the gravitational constant, $\sigma_0$ is the central velocity dispersion of the projected object, $\mu_0$ is the central surface brightness density and $r_{1/2}$ is the half light radius. 

In Figure \ref{fig:psats}(b) we plot the time evolution of the mass-to-light ratio for the satellites under study. It can be noticed that the mass-to-light ratio of the projected object, estimated using equation (\ref{eq:m2l}) remains relatively constant during the phase when the satellite is being slowly depopulated. After the satellite enters into the disruption phase, the mass-to-light ratio rises considerably reaching values of a hundred and more.
The values of the mass-to-light ratio of the satellites during the quasi-stable phase show the same kind of oscillations observed for the half light radius and the central velocity dispersion. During these oscillations the mass to light value varies over a wide range. Thus, depending on the time of observation a satellite might show very different values of the mass--to--light ratio. It is important to note that values much larger than the intrinsic mass to light ratio of the satellites can only be obtained during the quasi-stable phase of the evolution of the satellite. The simulated mass to light ratios are partially in good agreement with the observational data from real classical dSphs and UFDGs.

\subsection{Remnant density profiles}

\begin{figure}
\includegraphics[width=\linewidth,clip]{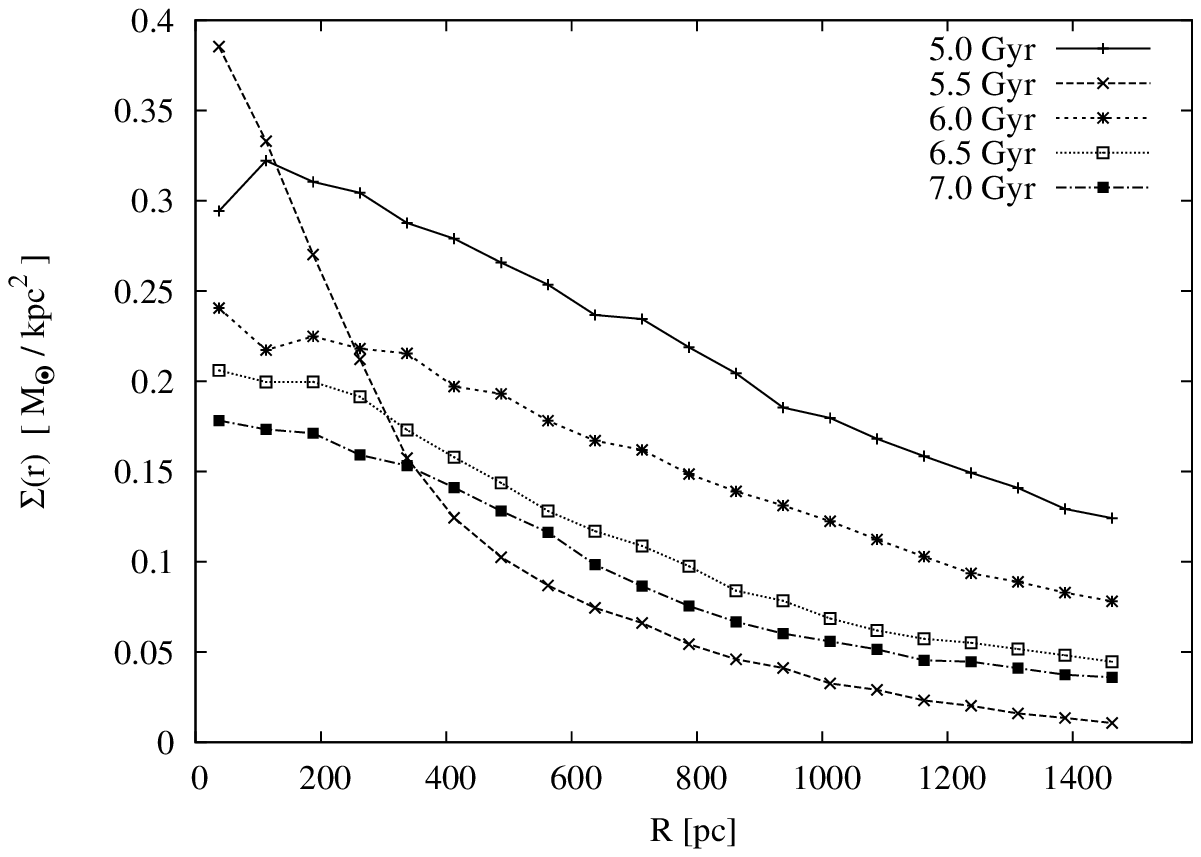}\\
\includegraphics[width=\linewidth,clip]{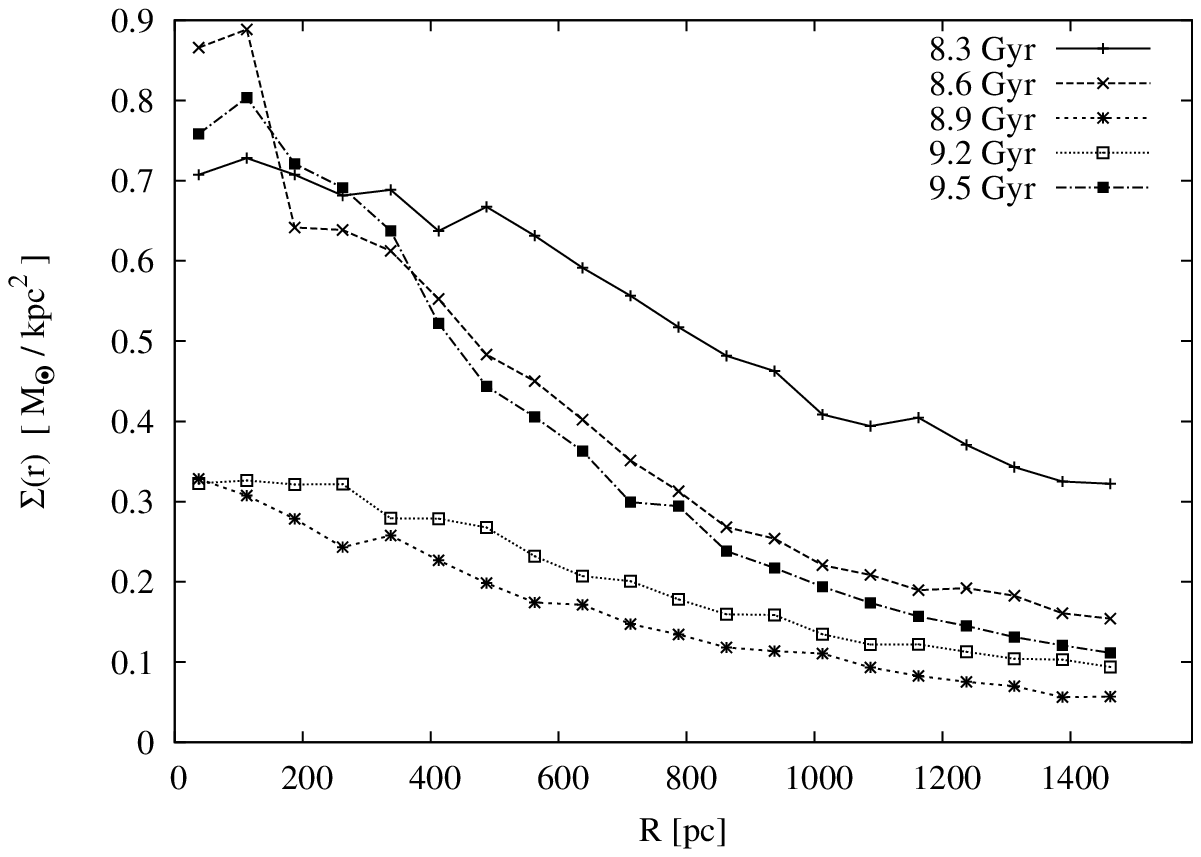}
\caption{Top: density profiles of Sat-1 on an orbit with an eccentricity of 0.73 and an apogalactic distance of  90 kpc. Bottom: density profiles of Sat-2 on an orbit with an eccentricity of 0.45 and an apogalactic distance of  30 kpc. The particular times of the profiles correspond to the quasi-stable time intervals respectively and are shown in the legends.}
\label{fig:dprofiles}
\end{figure}

In Figure \ref{fig:dprofiles} we show the projected surface density profiles of the remnant objects computed at several times during the quasi-stable phase of the satellites Sat-1 and Sat-2, respectively. The projected surface density profiles of the remnants are approximately exponential profiles with errors in the amplitude and $r_0$ below and around 5\% at the 68\% confidence level. Similar results are obtained in general for each satellite in the set of simulations. Therefore, it is clear that the remnant objects we are analysing are density distributions with a central density maximum. That is, they are physical density structures. 

\section{Comparison with dSphs and UFDGs populations}

Although the half light radius, the central surface brightness,  the central velocity dispersion along the line-of-sight and the mass to light ratio of the simulated satellites during their quasi-stable phases fall into the ranges of the corresponding quantities measured in real classical dSphs and UFDGs, from Figure \ref{fig:psats} it can be noticed that these simulated quantities do reproduce nearly but not exactly the observational quantities simultaneously, specially for less dense objects. In order to study further this result, we examine the scatter between the half light radius and the central surface brightness and between the half light radius and the central velocity dispersion of the simulated satellites and compare them with the available observations of real satellites.
 
\begin{figure*}
    \centering
    \includegraphics[width=0.9\linewidth,clip=yes]{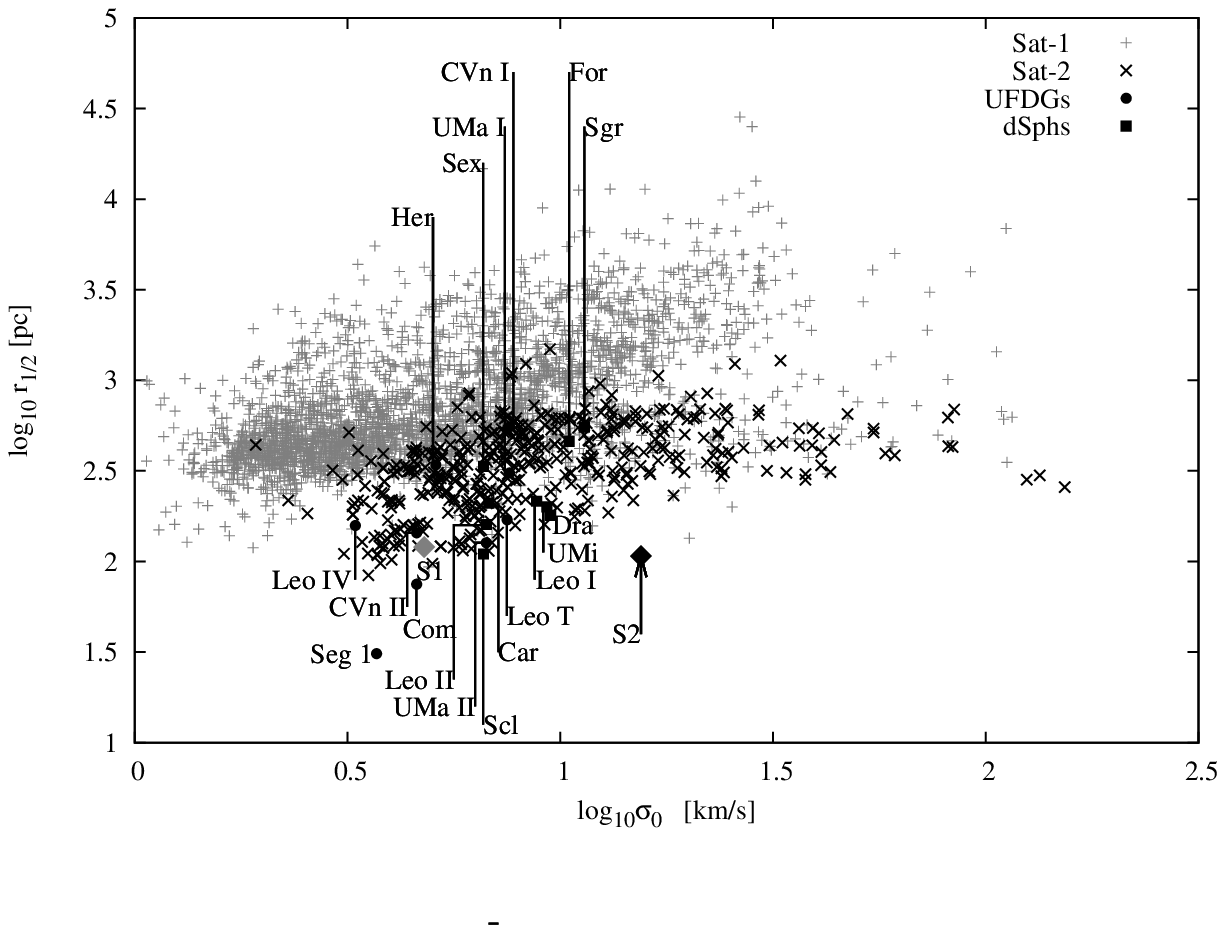}\\
    \includegraphics[width=0.9\linewidth,clip=yes]{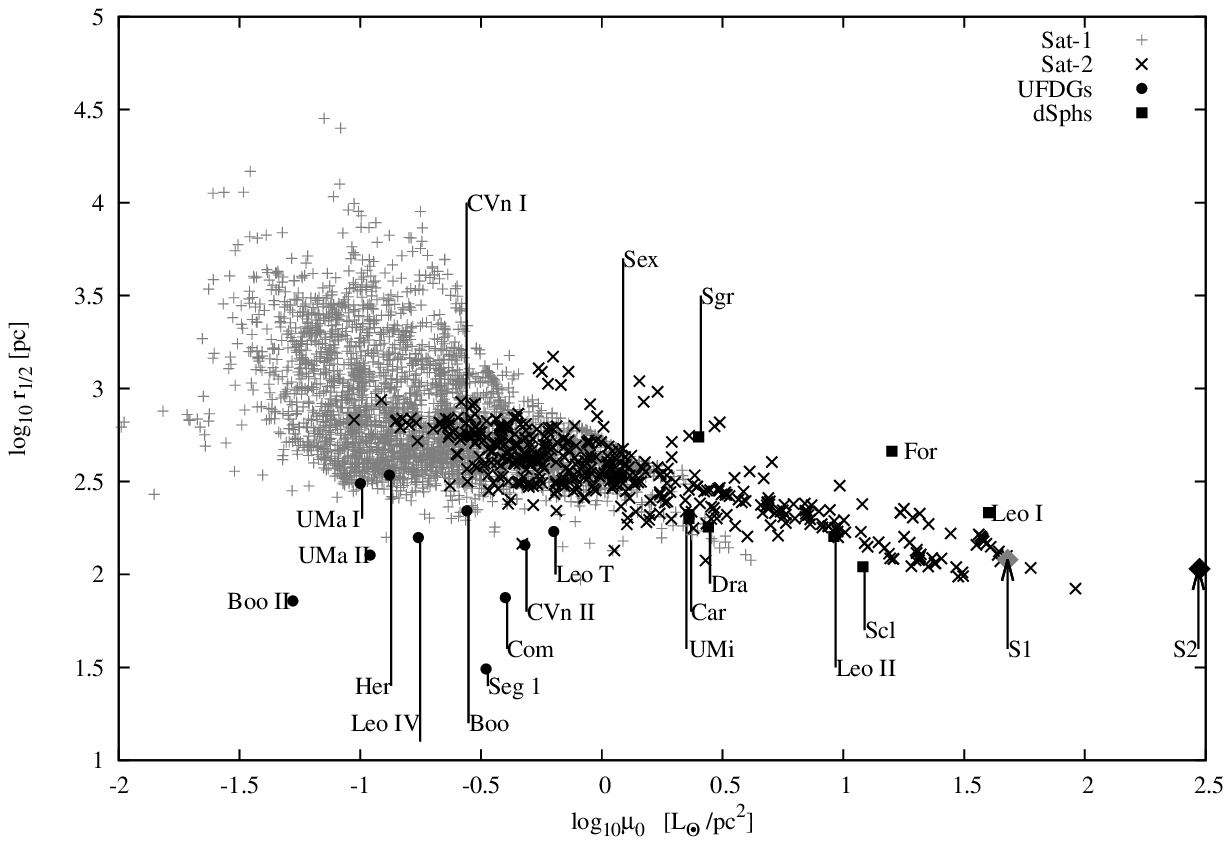}
   \caption{Half-light-radius versus central velocity dispersion (upper panel) and  the half-light-radius versus central surface brightness (lower panel) of the two satellites Sat-1 and Sat-2 under study, for all snapshots for all orbits in their quasi-stable phase, along with the corresponding quantities reported in the literature for real dSphs and UFDGs of the Milky Way. Large labeled diamonds show the corresponding quantities for the initial in-equilibrium Plummer objects Sat-1 (S1) and Sat-2 (S2). The observational data are given in Table \ref{tab:realdata}.} 
    \label{fig:realdata}
\end{figure*}

\begin{table}
\begin{centering}
\begin{tabular}{ccccc}
\hline
Galaxy & $r_{1/2}$  & $\sigma_0$   & $\mu_0$  & ref.\\
             &   (pc)       &(km/s)             & (mag /arcsec$^2$)  & \\
\hline
Sculptor  & 110 &  6.6    & 23.7 & 1\\
Fornax  & 460  & 10.5   & 23.4 & 1\\
Carina  &	210	& 6.8     & 25.5 & 1\\
Leo I	     & 215  & 8.8	     &  22.4 & 1	\\
Sextans &	 335	& 6.6     & 26.2  & 1 \\
Leo II & 160 & 6.7         & 24.0	& 1 \\
Ursa Minor &	200	& 9.3 & 25.5 & 1 \\
Draco         &   180	& 9.5 & 25.3 & 1 \\
Sagittarius  &	550	&11.4 &	25.4 & 1 \\
\\
Ursa Major II & 127 & 6.7 & 28.8 & 4\\
Leo T & 170 & 7.5 & 26.9 & 4,6\\
Ursa Major I & 308 & 7.6	& 28.9 & 4\\
Leo IV & 158 & 3.3 & 28.3 & 2,4 \\
Coma Berenices & 75 & 4.6 & 27.4 & 2,4\\
Canes Venatici II & 144 & 4.6 & 27.2 & 2,4\\
Canes Venatici I  & 554 & 7.6 & 27.8 & 4,7 \\
Hercules & 342	 & 5.1 & 28.6 & 2,4\\
Segue 1 & 31 & 3.7 & 27.6 & 8\\
Bootes II & 72	& -- & 29.6 & 3\\
Bootes & 220 & --	& 27.8 & 5\\
\hline
\end{tabular}
\end{centering}
\caption{Half light radius ($r_{1/2}$), central velocity dispersion along the line of sight ($\sigma_0$) and central surface brightness ($\mu_0$) of the nine classical dSphs and  eleven UFDGs. 
1: \citet{1998ARA&A..36..435M} and references therein, 2: \citet{2007ApJ...654..897B}, 
3: \citet{2007ApJ...662L..83W}, 4: \citet{2007ApJ...670..313S}, 5:  \citet{2006ApJ...647L.111B}, 
6: \citet{2007ApJ...656L..13I}, 7: \citet{2006ApJ...643L.103Z}, 8: \citet{2011ApJ...733...46S}.}
\label{tab:realdata}
\end{table}

In Figure \ref{fig:realdata} we plot both the half-mass-radius versus central velocity dispersion and  the half-light-radius versus central surface brightness of the two satellite types under study, in their quasi-stable phase, along with the corresponding quantities reported in the literature for the classical dSphs and UFDGs  of the Milky Way known up to day (see Table \ref{tab:realdata}). For the simulated data we have assumed a stellar mass-to-light ratio of 1, following \citet[]{1998ARA&A..36..435M}.
From the plots it can be observed that for both simulated satellites the data partially overlap the corresponding real galaxy data values. 
In particular, satellites like Sat-1 could be the progenitors of real galaxies like Sextans, Ursa Major I, Canes Venatici I and Hercules; and satellites like Sat-2 could be the progenitors of real galaxies like Carina, Sextans, Leo II, Ursa Minor, Draco and Canes Venatici I. 

It has to be emphasized that the simulated data plotted in Figure \ref{fig:realdata} correspond to satellites on different orbits and at different times. Therefore in order to confirm if real dSphs and UFDGs are remnants of initial objects like Sat-1 or Sat-2 it is necessary to perform additional simulations with larger numbers of particles and for particular orbits that could reproduce  the position and velocity of single real satellite galaxies and to prove that a single simulated satellite in a given orbit can reproduce simultaneously the observed quantities of a real galaxy.

\begin{figure}
    \centering
    (a)\\
    \includegraphics[width=\linewidth]{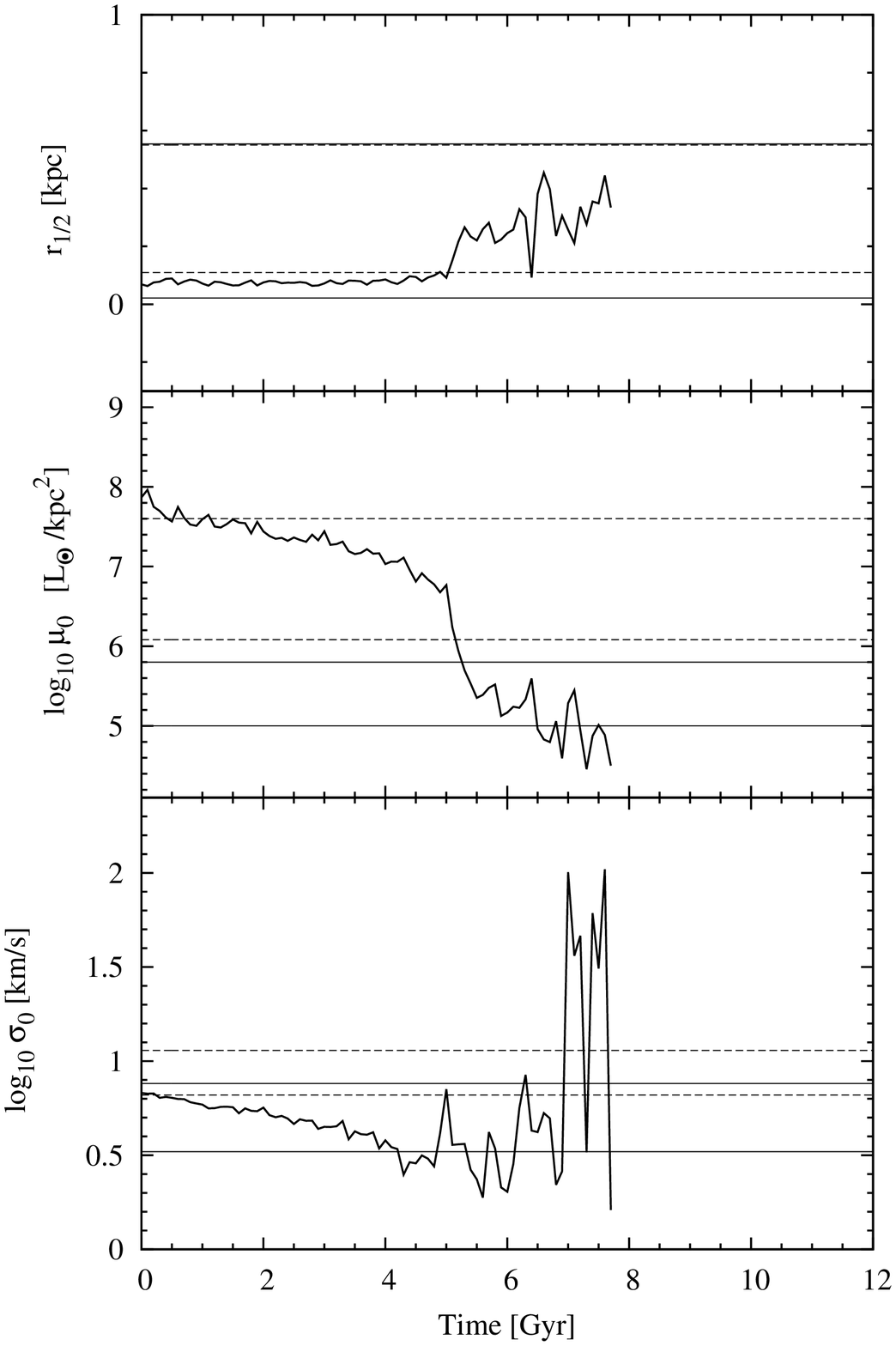}\\
    (b)\\
    \includegraphics[width=\linewidth]{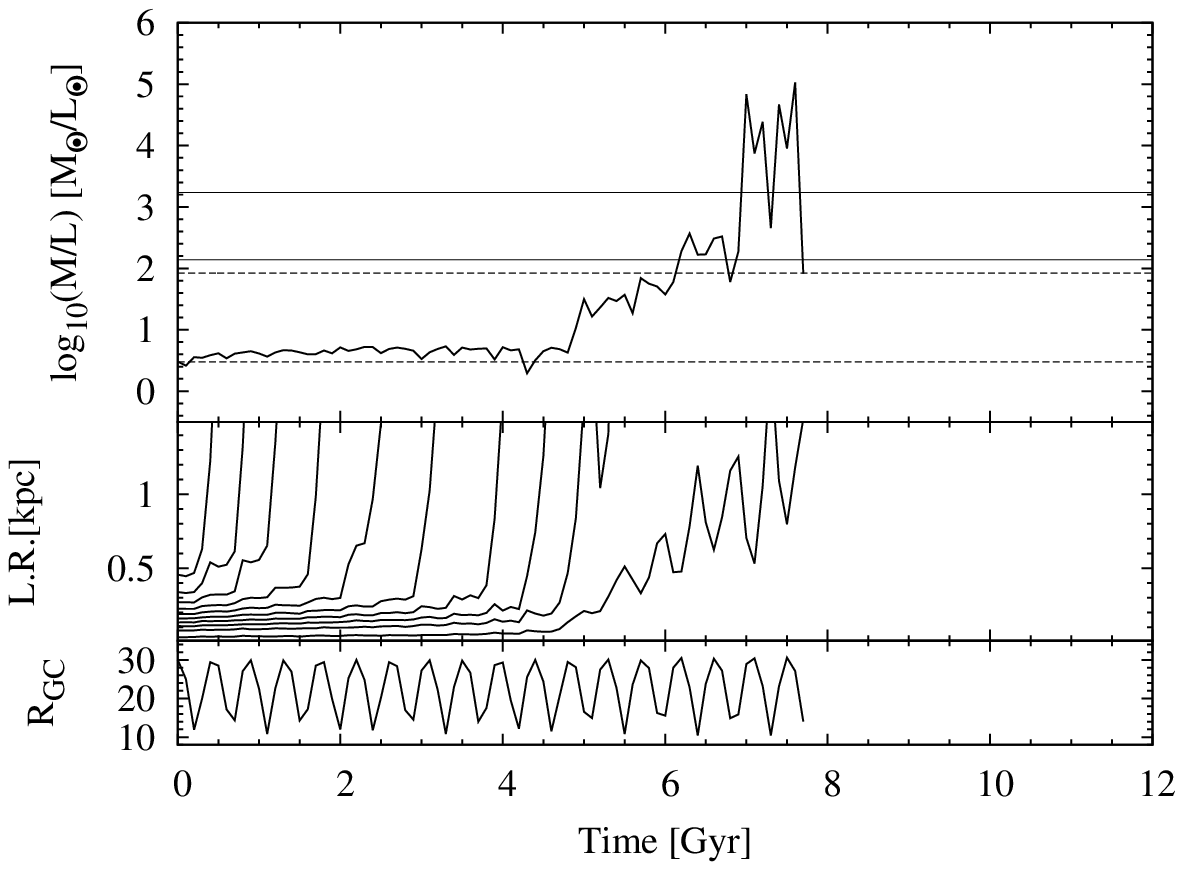}
   \caption{Observational quantities of a satellite (Sat-3) that exhibits a quasi-stable phase. (a) Half light radius, central surface brightness and central velocity dispersion along the line-of-sight. (b) Mass to light ratio, along with the corresponding Lagrange radii and Galactocentric distance of the centre of density of the objects. The horizontal lines show the range of observational data reported in the literature for classical dSph galaxies (dashed) and the new UFDGs (solid) of the Milky Way.} 
    \label{fig:sat-3}
\end{figure}

Since during the quasi-stable phase the simulated satellites increase their projected half light radius and decrease their central surface brightness, it is possible that an object with an initial Plummer radius smaller  than 0.3 kpc and total initial mass of $\approx 10^7$  $\mathrm{M}_\odot$ or lower can cover the region in the half mass radius -- central surface brightness plot corresponding to the UFDGs with low half light radii. To test this hypothesis we simulated the evolution of a satellite with an initial mass of $10^7$ \msun{} and a Plummer radius of 0.15 kpc (Sat-3) on a few orbits. We found that this progenitor on an orbit with an eccentricity of 0.45 and an apocentric distance of 30 kpc evolves into a quasi-equilibrium phase starting at $\approx 5$ Gyr and ending at  $\approx 7.5$ Gyr, as shown in Figure \ref{fig:sat-3}. During the time interval from  $\approx 6$ to  $\approx 7$ Gyr all the four properties under study of this satellite fit remarkably well into the ranges of the corresponding quantities reported for real UFDG´s of the Milky Way. Thus, initially dense objects like Sat-3 may be progenitors of some UFDGs of the Milky Way. Since this object is denser than Sat-1, the set of orbits that could lead to quasi-equilibrium remnants is expected to be smaller than the one for Sat-1. We found that Sat-3 progenitors on orbits with apocentric distances larger than 90 kpc can not evolve into the quasi-equilibrium phase within a Hubble time, which means that only UFDGs at distances smaller than 90 kpc could be associated with initial satellites like Sat-3 or with initially denser satellites. 

We have computed the fraction of time a simulated satellite in the quasi-stable phase looks
  similar to the real satellites of the Milky Way by computing the fraction of snapshots from
  all our models in the quasi-stable phase whose remnants present simulated
  properties that fall simultaneously into the corresponding observational
  ranges. For that we have taken all snapshots for which the mass-to-light ratio
  is  larger than 10 $M_{\odot}/L_{\odot}$ in Figure \ref{fig:realdata} and calculated which fraction of these snapshots have
  simultaneously a half-light-radius smaller than 500 pc, $\log (\sigma_0 [km/s])$ larger
  than 0.5 and  $log (\mu_0 [L_{\odot}/pc^2])$ larger than -1.3. We found that the satellites in the quasi-stable phase reproduce the observed satellite properties for $\approx 16\%$ of the orbit for Sat-1 progenitors, $\approx 66\%$ for Sat-2 progenitors and $\approx 42\%$ for the single Sat-3 progenitor.

\section{Conclusions}
The Newtonian simulations performed here show that it is possible to obtain disrupted remnants of initially spherical bound objects, that survive as quasi-stable systems for times longer than 1 Gyr, confirming and extending the results from \citet{1997NewA....2..139K}, but for a satellite with more particles and for a three-component galaxy model, that resembles the Milky Way. From a large set of numerical simulations of the long term time evolution of two satellites with different initial density orbiting the Milky Way in a variety of orbits two maps of the orbital conditions that lead to quasi-equilibrium objects were constructed. We have found that the disrupted  satellites show high mass-to-light ratios, which have similar values to those inferred for some classical dSphs and UFDGs of the Milky Way and that the whole set of simulations reproduce the relationships between the half light radius and central velocity dispersion along the line of sight and between the half light radius and central surface brightness.
Thus, within the Newtonian framework, objects like Sat-1, Sat-2 and Sat-3 do show remarkable similarities to the real satellites. 

The presented Newtonian models do not perfectly represent the observed satellites in terms of the surface densities, radii and M/L values at a given time.  However, within the model where TDG satellites are born $\approx 8$ Gyr ago within a single event and are now in the quasi-stable phase, 10\% - 66\% could be influenced at the current time by tidal effects which enlarge the dynamical M/L substantially.

Further simulations with larger mass resolutions are needed to improve our understanding of whether UFDGs and dSph satellites are ancient tidal dwarf galaxies in Newtonian dynamics.

The results suggest that a minor fraction of the observed satellites could plausibly be galaxies without dark matter that have true M/L ratios much lower those those measured. The inflated M/L ratios arise because they are observed at the right time, along the right orbit and  during the quasi-equilibrium phase of their evolution. This to be a viable explanation in Newtonian dynamics for the high M/L ratios observed in all satellites, the satellite would need to be preferentially on certain orbit and observed at certain times. This {\it could} arise within the TDG scenario if {\it  all} satellites are created at the same time along a few specific orbits that are particularly susceptible to the quasi-equilibrium phase. \citet{2010ApJ...722..248M} have however shown that tidaly-influenced dwarf galaxies without dark matter are well understandable in Milgromian dynamics.

It is to be emphasized that the computations reported here have been performed in Newtonian dynamics. If the dSphs and UFDGs of the MW are ancient dark-matter-free tidal dwarf galaxies (TDG), as is strongly 
suggested to be the case by their correlation in phase-space \citep{2010A&A...523A..32K,  2011A&A...532A.118P}, then the work presented here would need to be repeated in a non-Newtonian framework to be logically consistent with the implications which the TDG scenario would have.  Indeed, an approach in MOND to understand the internal dynamics of dSphs and UFDGs has been shown to be very promising \citep{2010ApJ...722..248M}.

\section*{Acknowledgements}
 R.A.C. and K.P.R. acknowledge financial support  from the \textit{
   Convocatoria Nacional de Investigaci\'on 2008  de la Direcci\'on de
   Investigaciones de  la Sede Bogot\'a de la Universidad Nacional de
   Colombia}.

\bibliography{my_citations}

\label{lastpage}

\end{document}